\documentclass[
onecolumn,
superscriptaddress,
 amsmath, amssymb,
aps,
dblfloatfix,
nobalancelastpage,
notitlepage
]{revtex4-1}
\usepackage{graphicx}
\usepackage{dcolumn}
\usepackage{bm}
\usepackage{siunitx}
\usepackage{hyperref}
\usepackage{braket}
\usepackage{epstopdf}
\usepackage{filecontents}
\hypersetup{colorlinks=true, citecolor=blue, urlcolor=blue, linkcolor=blue}

\makeatletter
\def\maketitle{
\@author@finish
\title@column\titleblock@produce
\suppressfloats[t]}
\makeatother

\begin{document}

\preprint{APS/123-QED}

\title{Emergent intersubband-plasmon-polaritons of Dirac electrons under one-dimensional superlattices
}

\author{Minwoo Jung}
\email{mj397@cornell.edu}
\affiliation{Department of Physics, Cornell University, Ithaca, New York, 14853, USA}
\author{Gennady Shvets}
\email{gs656@cornell.edu}
\affiliation{School of Applied and Engineering Physics, Cornell University, Ithaca, New York 14853, USA}

\date{\today}

\begin{abstract}
Artifical superlattice (SL) potentials have been employed extensively for band structure engineering of two-dimensional (2D) Dirac electron gas in graphene \cite{BandEngineer1,BandEngineer2,BandEngineer3,nat_nano_SiO2,BandEngineer5}. While such engineered electronic band structures can modify optical or plasmonic properties of graphene, an emergent polaritonic behavior beyond weak perturbative effects (e.g. anisotropic Drude weights \cite{Brey_PRL}) has not been reported. Here, we show that an extreme modulation of one-dimensional (1D) SL potentials in monolayer graphene deforms the underlying Dirac band dispersion and introduces ladder-like energy levels near the Fermi surface, which result in emergent intersubband polaritonic responses in optical conductivity. In our proposed system, hBN-encapsulated graphene is placed on top of a 1D periodic metagate. In addition, a backgate placed beneath the metagate is used as the second gate, further modulating carrier density on regions in graphene that are not directly screened by the metagate. With a strong carrier density modulation, graphene electrons experience an array of deep potential wells, and at large enough momenta perpendicular to the modulation direction, Dirac electrons are waveguided via total internal reflections. These waveguided modes appear as flat subbands with nearly equispaced energy levels. As a result, there arise hybrid intersubband-polaritons with ultra-strong coupling in plasmonic dispersions. Our study opens up an avenue for exploring emergent polaritons in two-dimensional materials with gate-tunable electronic band structures.
\end{abstract}

\maketitle
Polaritons in 2D semiconductor materials (e.g. plasmon polaritons in graphene \cite{hBN1, hBN3} or exciton-trion polaritons in transitional metal dichalcogenides \cite{ETP1, ETP2}) promise an ideal platform for novel opto-electronic devices, supported by various ways to control the carrier densities in these 2D materials. Photo-carrier injection via high-intensity pulses is useful to probe ultrafast transient responses of such polaritons \cite{photo_mod1, photo_mod2, photo_mod3}, and capacitive field-effect gating is exploited for active control over steady-state responses. While a uniform gate can tune overall polariton dispersions due to carrier density dependence of Drude weight in graphene \cite{hBN1, hBN3} and exciton/trion oscillator strengths in transitional metal dichalcogenides \cite{ETP_mod}, more exotic electro-optic controls can be achieved by using a metagate--a gating structure with spatially varying capacitance. Especially in graphene plasmonics, spatial modulation of carrier densities via metagate-tuning has been considered for various applications such as local phase modulation \cite{hBN4}, topological phase switching \cite{GP_mod2}, and Bloch polariton steering \cite{GP_mod3}.

\begin{figure}[b]
    \includegraphics[width=0.95\columnwidth]{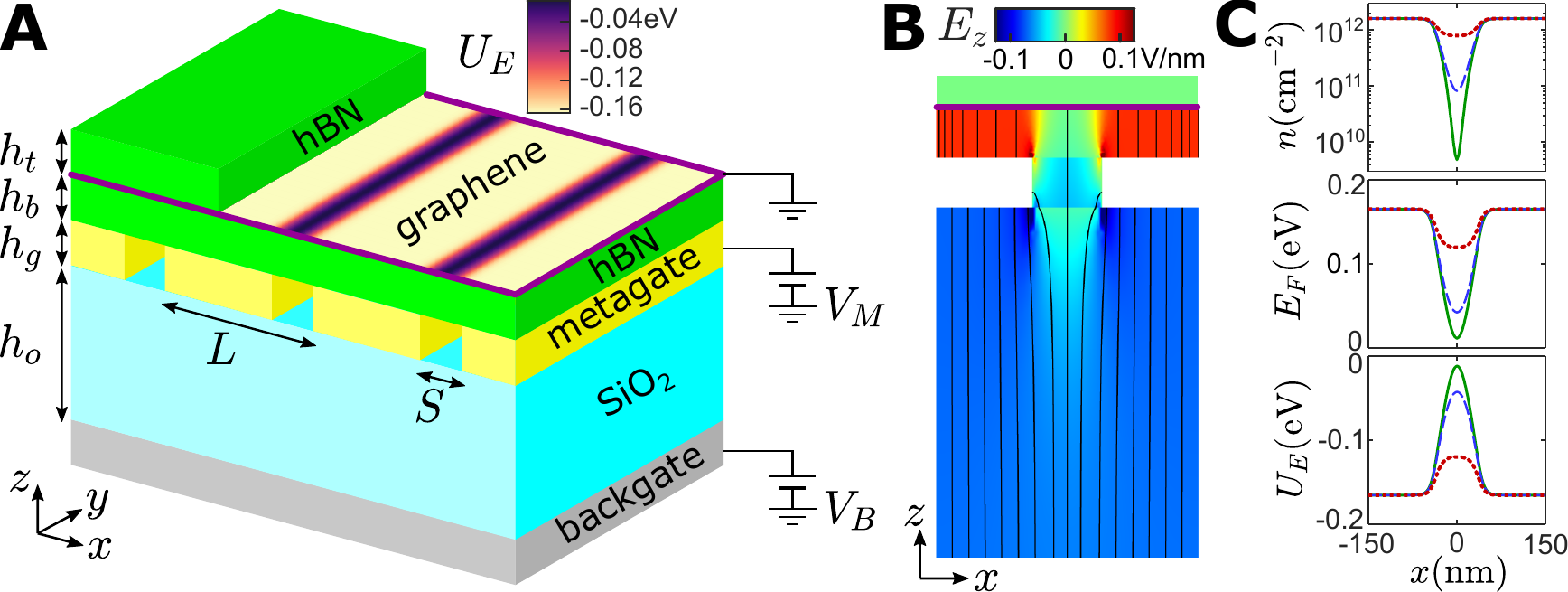}
    \caption{SL engineering with double(metagate/backgate)-gating. \textbf{a} Schematic of the device design; the inset on the exposed graphene plane is an example of a SL potential ($h_t=5$nm, $h_b=10$nm, $h_g=10$nm, $h_0=150$nm, $L=300$nm, $S=80$nm, $V_M=1$V and $V_B=-9$V). \textbf{b} Electrostatic simulation showing the electric field along the direction perpendicular to graphene sheet, $E_z$, and the electric field lines in $x-z$ plane; the role of the backgate is to further modulate the carrier density locally on the region that is not screened by the metagate. \textbf{c} The doped carrier density $n(x)$ (top), the corresponding Fermi level $E_F(x)$ (middle), and the SL potential $U_E(x)$ (bottom) for different backgate voltages ($V_M=1$V is for all three cases); dotted red: $V_B=12$V, dashed blue: $-6$V, and solid green: $-9$V.}
    \label{fig.1.}
\end{figure}

Owing to quantum capacitance effect \cite{QCG}, spatially modulated carrier densities $n(\mathbf{r})$ under electrochemical equilibrium gives rise to SL electric potentials $U_E(\mathbf{r})$ for Dirac electrons, given as
\begin{equation}
\label{eq1}
\hbar v_F \sqrt{\pi n(\mathbf{r})} + U_E(\mathbf{r}) = \mu_0,
\end{equation}
where $v_F$ is the Fermi velocity of Dirac electrons and $\mu_0$ is the electrochemical potential. Under a periodic SL potential, the conical Dirac dispersion deforms into minibands or subbands \cite{BandEngineer1,BandEngineer2,BandEngineer3,nat_nano_SiO2,BandEngineer5}. These SL-induced subbands has been probed through emergent electronic transport properties \cite{nat_nano_SiO2,BandEngineer5}, but an emergent optical or polaritonic phenomenon directly stemming from this band structure deformation hasn't been reported. Even though the above-mentioned works \cite{hBN4, GP_mod2, GP_mod3} utilized spatial modulation of $n(\mathbf{r})$, their phenomena are fully captured at the level of a simple optical conductivity model with locally varying Drude weights computed from the conical Dirac dispersion before SL-induced deformation. We note that a recent study \cite{Brey_PRL} investigated plasmonic responses in graphene under a strict form of SL potentials $U_E(x) \propto \cos{(G_0 x)}$, and reported anisotropic plasmonic dispersions and increased plasmon damping due to subband transitions. While such marginal changes in graphene plasmons are certianly direct consequences of SL engineering, these effects don't necessarily introduce a new type of polaritonic excitation.

In this study, we show that a 1D SL potential in a 2D Dirac electron gas leads to coherent intersubband resonances in the optical conductivity. Along with plasmon-polaritons, these newly emergent intersubband-polaritons form hybrid intersubband-plasmon-polaritons (HIPPs). Each SL unit cell carries a square potential well deep enough to host multiple tightly-confined bound states with negligible coupling to bound states in adjacent unit cells, thereby forming flat subbands in the deformed band structure. Notably, the massless nature of Dirac electrons plays a pivotal role in the ultrastrong resonant enhancement of oscillator strengths of the intersubband transition (ISBT) at certain quantized frequencies, as the subbands are given at ladder-like equispaced energy levels unlike the quadratic energy levels of massive particles in a square potential well. Also, in our proposed system, the HIPP dispersions can be systemically controlled via the combination of a metagate and a backgate, where the metagate controls the underlying plasmon-polariton dispersions and the backgate controls the intersubband resonance frequencies and the Rabi-splitting strengths. Our results demonstrates that SL engineering in 2D materials, a versatile technique to study exotic electronic transport properties, can be used for the search of novel polaritonic materials as well.

Figure \ref{fig.1.}\textbf{a} depicts the geometry of the considered device platform. The combination of a patterned metagate and an additional normal gate has been used in several other works \cite{BandEngineer5, GP_mod2} to attain better tunability of graphene carrier densities. We put the flat gate as a backgate beneath
the metagate, so that the backgate is used to control the carrier densities only at a particular region on graphene that is not screened by the metagate. In this way, if the duty cycle of air gaps in the metagate grating is not so large ($S/L<50\%$, roughly speaking), the baseline Drude weight ($\propto E_F=\hbar v_F \sqrt{\pi n}$) of graphene optical conductivity is mostly determined by the metagate voltage $V_M$, and the SL modulation depth is fine-tuned with the backgate voltage $V_B$. The electrostatic simulation results shown in Fig. \ref{fig.1.}\textbf{b}-\textbf{c} are carried out by an iterative solver in COMSOL Multiphysics to match the nonlinear boundary condition given in Eq. (\ref{eq1}) \cite{Minwoo_PRL}, and we consider the density-dependent renormalization of the Fermi velocity: $v_F(n)=[0.85+0.035\ln(n_0/n)]\times 10^6$m$/$s, where $n_0=10^{15} $cm$^{-2}$ \cite{novoselov_pnas13}. Figure \ref{fig.1.}\textbf{b} clearly shows that the electric field lines from the backgate is mostly screened by the metagate, but a few of them would penetrate through the air gap and reach to graphene. As a result, the doped carrier density in the region above the metallic grating is flat and controlled by $V_M$, whereas the region above the air gaps can even be almost depleted when a large enough negative voltage $V_B$ is applied to the backgate (Fig. \ref{fig.1.}\textbf{c}). Since we consider graphene to be grounded $\mu_0=0$, the SL potential is given by simply negating the Fermi level $U_E(x)=-E_F(x)$. The resulting shape of $U_E(x)$ can be viewed as a periodic array of wide square potential wells with narrower separating walls, where the height of separating walls is directly controlled by $V_B$.

\begin{figure}[b]
    \includegraphics[width=0.95\columnwidth]{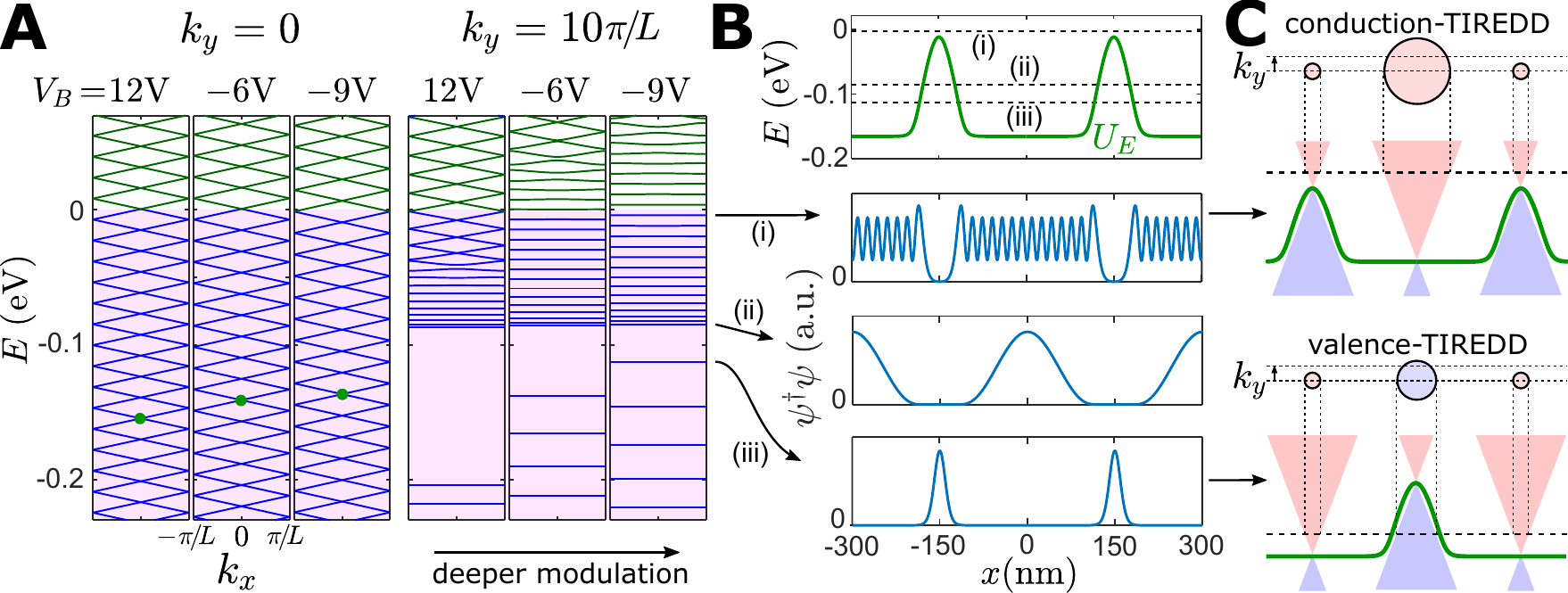}
    \caption{Ladder-like flat subbands of Dirac electrons via TIREDD. \textbf{a} Left: Subbands with $k_y=0$ appear as simply folded conical dispersions due to Klein tunneling. Filled bands ($E<\mu_0=0$) are colored in blue, and unoccupied bands are in green. The red dots are Dirac crossing points. Right: With a sufficiently large value of $k_y$ and a strongly modulated SL potential, several subbands around the Fermi surface appear as flat bands along $k_x$. Not only that, the level spacing of these flat subbands is almost uniform due to the nature of Dirac electrons that have linear dispersion. \textbf{b} The energy diagram on top shows the energy levels of several example bound states--(i) a mode right below the Fermi surface, (ii) the lowest energy bound state formed by conduction band TIREDD, and (iii) the highest energy bound state formed by valence band TIREDD--relative to the SL potential landscape ($V_B=-9$V). The following graphs are the wavelength amplitudes $\psi^\dagger\psi$ of these three example bound states. Each potential well in each unit cell confines the bound states so tightly that the tunneling into the adjacent unit cells is vanishing, which explains the appearance of the flat subbands. \textbf{c} Schematics for the mechanism of TIREDD for the waveguiding of conduction band electrons (top) and of valence band electrons (bottom). Each circle represent the iso-energy contour of a Dirac cone in the momentum space, and its radius $k_r(x)=|E-U_E(x)|/\hbar v_F$ is determined by the difference of the eigenenergy level (dashed lines) and the SL potential (green lines) at a local position. Electron waves can propagate along $x$-axis if $k_y$ is less than $k_r$, but it evanescently decays along $x$-axis if $k_y$ is greater than $k_r$. Thus, when an electron travels from a domain with $k_y<k_r$ to another domain with $k_y>k_r$, there occurs TIREDD. These TIREDD-based bound states can be formed both in potential wells (top; for conduction band electron) and in potential barriers (bottom; for valence band electron).}
    \label{fig.2.}
\end{figure}

The electronic subband structures for Dirac electrons in 1D SL potentials periodic along the $x$-axis are calculated by solving Dirac Hamiltonian equation $\left(v_F\nu p_x\sigma_x+v_F p_y\sigma_y+U_E\mathcal{I}\right)\ket{\mathbf{k};j}=E_{\mathbf{k};j}\ket{\mathbf{k};j}$ with the Bloch ansatz $\braket{\mathbf{r}|\mathbf{k};j}=e^{i\mathbf{k}\cdot\mathbf{r}}\psi_{\mathbf{k};j}(x)$
\begin{equation}
\label{eq2}
\begin{bmatrix}
U_E(x) & \hbar v_F \nu (k_x-i\partial_x)-i\hbar v_F k_y\\
\hbar v_F \nu (k_x-i\partial_x)+i\hbar v_F k_y & U_E(x)
\end{bmatrix}\psi_{\mathbf{k};j}(x) = E_{\mathbf{k};j}\psi_{\mathbf{k};j}(x),
\end{equation}
where $\nu$ is the valley index ($+1$ for $\mathbf{K}$-valley and $-1$ for $\mathbf{K}^\prime$-valley), $j\in\mathbb{Z}$ is the subband index, and the Bloch wavefunction satisfies the periodicity condition $\psi_{\mathbf{k};j}(x)=\psi_{\mathbf{k};j}(x+L)$. Here, $\mathbf{k}=k_x \hat{\mathbf{x}} + k_y\hat{\mathbf{y}}$ is the electron wavevector, where $k_x$ is the Bloch wavenumber parallel to the periodic direction and $k_y$ is the wavenumber perpendicular to the SL modulation. The same band structure is repeated for each spin subspace. With vanishing $k_y=0$, the conical linear dispersion is exactly preserved even in the presence of strongly modulated $U_E(x)$ (Fig. \ref{fig.2.}\textbf{a} left) due to Klein tunneling \cite{Klein1, Klein2}. At sufficiently large $k_y$, in contrast, there occur several flat subbands around and below the Fermi surface (Fig. \ref{fig.2.}\textbf{a} right), as the bound states are formed inside the potential wells via the total internal reflection of electrons with Dirac dispersion (TIREDD). As clearly shown in Fig. \ref{fig.2.}\textbf{b}, these bound states are tightly confined within each unit cell and their wavefunction amplitudes $\psi^\dagger\psi$ decay rapidly to zero across the separating potential walls so that Dirac electrons are waveguided along $y$-axis with negligible tunneling across the adjacent wells. The negligible coupling between the bound states in the adjacent unit cells manifests as flat subbands in the band diagrams.

The TIREDD condition is met when a Dirac electron at a given energy $E$ travels with a nonzero $k_y$ from a domain, where the propagation along $x$-axis is allowed, to another domain, where the propagation is not allowed. At a local position along the SL potential, the iso-energy contour of the conial Dirac dispersion is given as a circle in the momentum space with a radius of $k_r(x)=|E-U_E(x)|/\hbar v_F$. Therefore, the local wavenumber along $x$-axis $\sqrt{k_r(x)^2-k_y^2}$ becomes imaginary if $k_y$ exceeds $k_r(x)$, and the electron wavefunction decays exponentially along $x$-axis as $\psi\propto e^{-\alpha x}$, where the decaying factor is given as $\alpha = \sqrt{k_y^2 - k_r^2}$. It is evident that this evanescent decay along $x$-axis is not possible with $k_y=0$, which, along with the linear nature of Dirac dispersion, leads to Klein tunneling as discussed earlier. Notably, as illustrated in Fig. \ref{fig.2.}\textbf{c}, TIREDD occurs not only when the electron travels from a lower potential domain to a higher potential domain--a more familiar picture--but it also happens in the opposite case when the electron travels from a higher potential domain to a lower potential domain. The latter case corresponds to TIREDD of an antiparticle in the high-energy physics language, or to TIREDD of the valence band electron in graphene. As a result, bound states out of valence band electrons are formed around the potential barriers, see the state (iii) in Fig. \ref{fig.2.}\textbf{b}. Each flat subband in the band diagram in Fig. \ref{fig.2.}\textbf{a} corresponds to a TIREDD-based bound state, and the flatness depends on whether the width of decaying barriers, the regions satisfying $k_y>k_r(x)$, is sufficiently thicker than the inverse of decaying factor $\alpha$. We note that the mean free path of electrons in hBN-encapsulated graphene can be as high as $1\mu$m already at the room temperature and near $10\mu$m at cryogenic temperatures $T<100$K \cite{high_mobility}. Thus, we expect that the TIREDD condition can be realized in the realistic experiments, since the mean free path in high-quality samples is much longer than the width of the SL potential wells.

\begin{figure}[b]
    \includegraphics[width=0.45\columnwidth]{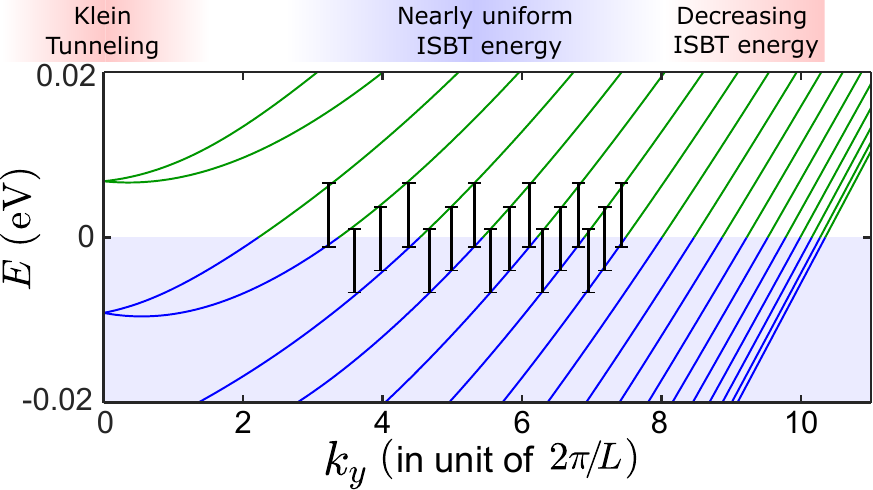}
    \caption{The massless dispersion of Dirac electrons makes the ISBT energy to be nearly uniform over a broad region Fermi surface. The electronic subband structure is shown along $k_y$ at a fixed $k_x=0$ ($V_B=-9V$ case). Each vertical black bar is given as a guide to eyes for denoting a vertical transition ($\Delta j = 1$) from an occupied state below the Fermi surface to a state above the Fermi surface, and all bars have have the same length. In this case, the ISBT energy appears as very uniform roughly within $6\pi/L<|k_y|<15\pi/L$, which is almost 45\% of the whole area of the Fermi surface $|k_y|<20\pi/L$.}
    \label{fig.3.}
\end{figure}

Another notable feature of these TIREDD-based flat subbands is the ladder-like energy level spacing around the Fermi surface. With a rough approximation (for more detailed analysis, see the Supplementary Information), the bound state energy levels of Dirac electrons in a square potential well (for now, let's consider the conduction band TIREDD only) are given as $E_j\sim\hbar v_F \sqrt{(\pi j/W)^2 + k_y^2}+U_0$, where $W$ is the width of potential well, $U_0$ is the potential inside the well, and $j$ is the band index. Even though this expression is not completely linear in $j$, it quickly approaches to the asymptotic linear relation $E_{j+1}-E_j\sim\pi\hbar v_F /W$ when $(E_j-U_0)$ is only twice greater than $\hbar v_F k_y$. Thus, the ISBT energy is maintained as nearly uniform over a substantial portion of the Fermi surface (see Fig. \ref{fig.3.}), which resonantly enhances the oscillator strength of the ISBT at  certain quantized frequencies given at integer multiples of $\pi v_F/W$. In Fig. \ref{fig.3.}, six or seven bands are altogether contributing to the ISBT nearly at the same frequency.

\begin{figure}[b]
    \includegraphics[width=0.85\columnwidth]{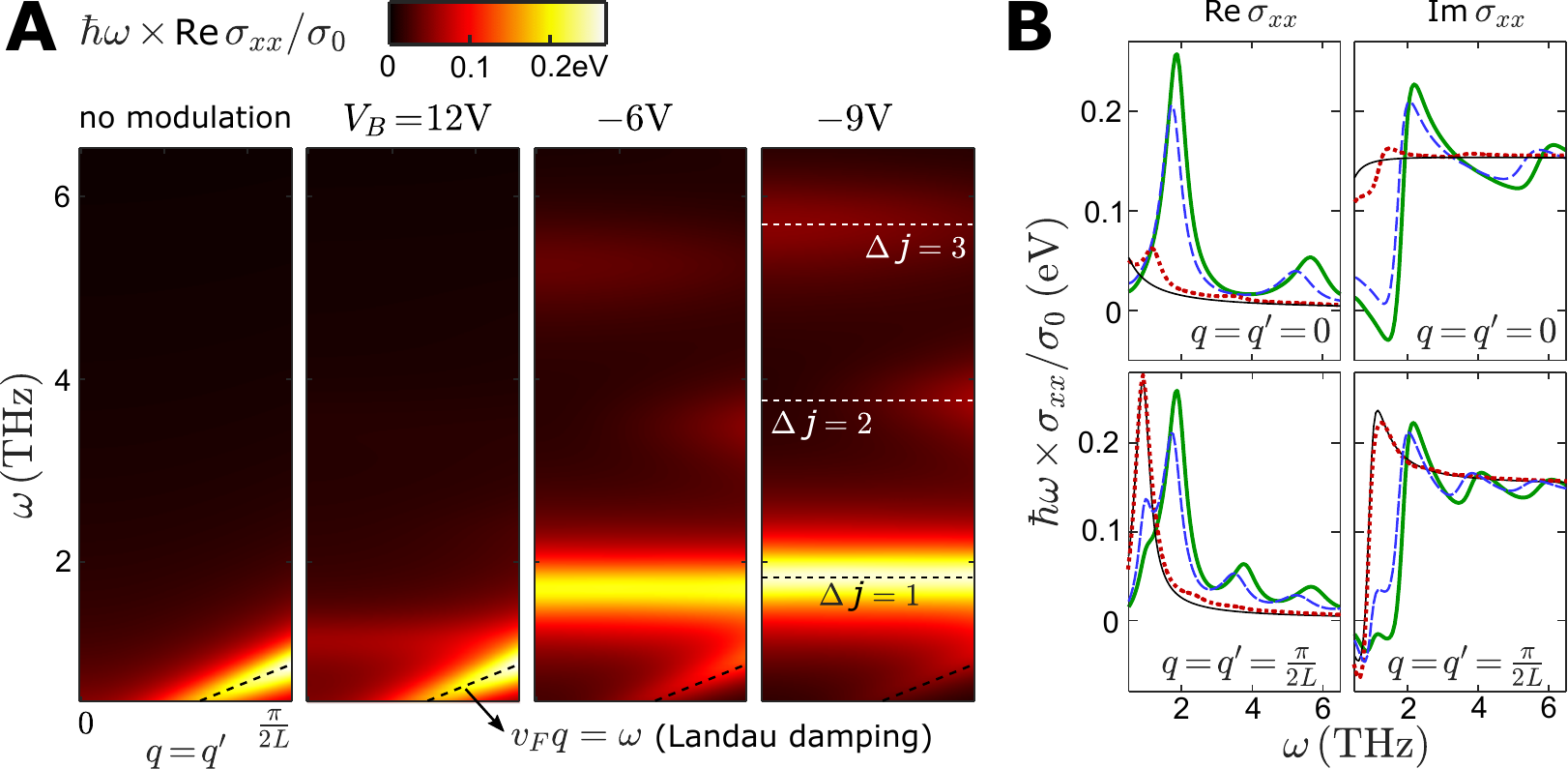}
    \caption{Graphene optical conductivity and ISBT resonances. \textbf{a} Real part of the conductivity Re$[\sigma_{xx}(\mathbf{q}=q\hat{\mathbf{x}},\mathbf{q}^\prime=q^\prime\hat{\mathbf{x}};\omega)]$ calculated for $q=q^\prime$; in order to visualize the features at higher frequencies better, we plotted $\omega\times\sigma_{xx}$. With strongly modulated SL potentials ($V_B=-6$V, $-9$V), there appear several Lorenzian peaks that correspond to the ISBT. $\Delta j$ refers to the difference between subband indices of two bands involved in the ISBT. As a comparison, the conductivity for uniformly doped ($E_F=0.15$eV) graphene is also provided. \textbf{b} Both real (left) and imaginary (right) part of the conductivity calculated at $q=q^\prime=0$ (top) and $q=q^\prime=\frac{\pi}{2L}$ (bottom); solid thin black: no modulation ($E_F=0.15$eV), dotted red: $V_B=12$V, dashed blue: $-6$V, and solid thick green: $-9$V.}
    \label{fig.4.}
\end{figure}

Ultrastrong ISBT optical responses manifest as resonant features in graphene optical conductivity, which is calculated by the Kubo formula under random phase approximation (RPA) \cite{Brey_PRL, RPA}, given as
\begin{equation}
\label{eq3}
\frac{\sigma_{xx}(\mathbf{q},\mathbf{q}^\prime;\omega)}{\sigma_0} = i\pi g_sg_\nu\hbar^2 v_F^2\int \frac{d^2\mathbf{k}}{(2\pi)^2}\int \frac{d^2\mathbf{k}^\prime}{(2\pi)^2} \sum_{j,j^\prime} \frac{f(E_{\mathbf{k};j})-f(E_{\mathbf{k}^\prime;j^\prime})}{E_{\mathbf{k};j}-E_{\mathbf{k}^\prime;j^\prime}}\frac{\bra{\mathbf{k};j}e^{-i\mathbf{q}\cdot\mathbf{r}}\sigma_x\ket{\mathbf{k}^\prime;j^\prime}\bra{\mathbf{k}^\prime;j^\prime}e^{i\mathbf{q}^\prime\cdot\mathbf{r}}\sigma_x\ket{\mathbf{k};j}}{\hbar(\omega + i\gamma)+E_{\mathbf{k};j}-E_{\mathbf{k}^\prime;j^\prime}},
\end{equation}
where $\sigma_0=2e^2/h$ is the conductance quantum, $g_s=2$ and $g_\nu=2$ are the spin and valley degeneracy, $f(E)=1/\left[\exp(-E/k_BT)+1\right]$ is the Fermi-Dirac distribution ($\mu_0=0$), and each 2D momentum integration is done as $\int d^2\mathbf{k}=\int_{-\pi/L}^{\pi/L}dk_x\int_{-\infty}^{\infty}dk_y$. To avoid any confusion for notations, we use $q$ to denote the momentum of the optical field, whereas we have used $k$ to denote the momentum of Dirac electrons. Since our system is periodic in $x$-axis, the conductivity is vanishing except when $\mathbf{q}-\mathbf{q}^\prime$ is an integer multiple of $G_0\hat{\mathbf{x}}$ ($G_0=\frac{2\pi}{L}$); therefore, the surface current density response is given as $K_x(\mathbf{q})=\sum_{\mathbf{q}^\prime\in\{\mathbf{q}+mG_0\hat{\mathbf{x}}|m\in\mathbb{Z}\}}\sigma_{xx}(\mathbf{q},\mathbf{q}^\prime)E_x(\mathbf{q}^\prime)$. In this work, we only consider the plasmonic excitations along $x-$axis ($\mathbf{q}=q\hat{\mathbf{x}}$, $\mathbf{q}^\prime=q^\prime\hat{\mathbf{x}}$) with transverse-magnetic polarization ($B_x=E_y=0$) \cite{TETM}. For the results plotted in Fig. \ref{fig.4.}, we assumed a cryogenic temperature of $T=60$K and a plasmonic scattering rate (Drude loss) of $\gamma=2\pi\times0.2$THz, which are well within the experimentally attainable ranges \cite{basov_nature_2018}. We calculated the conductivity $\sigma_{xx}$ for each of three SL potentials given in Fig. \ref{fig.1.}\textbf{c} and also for an unmodulated graphene with uniform doping of $E_F=0.15$eV as a comparison. Even in the no-modulation case, a resonant behavior is found along a linear line $v_Fq=\omega$. This corresponds to the intraband transitions that occur when the phase velocity of the optical field matches with the Fermi velocity of Dirac electrons. This velocity-matching effect becomes one of the key distinguishing factors between the simple Drude conductivity model and the nonlocal RPA conductivity model in uniformly doped graphene \cite{koppens_science_17}. When the SL modulation depth is moderate ($V_B=12$V case), only the first ($\Delta j=1$; $\Delta j$ refers to the difference between two subband indices) ISBT resonance appears faintly, while the conductivity at higher frequency is almost the same to the no-modulation case. With much stronger modulation depths, however, the first ISBT resonance peak becomes the most prominent feature and the higher-order ($\Delta j=2,3,...$) ISBT resonance peaks also become visible. 

\begin{figure}
    \includegraphics[width=0.8\columnwidth]{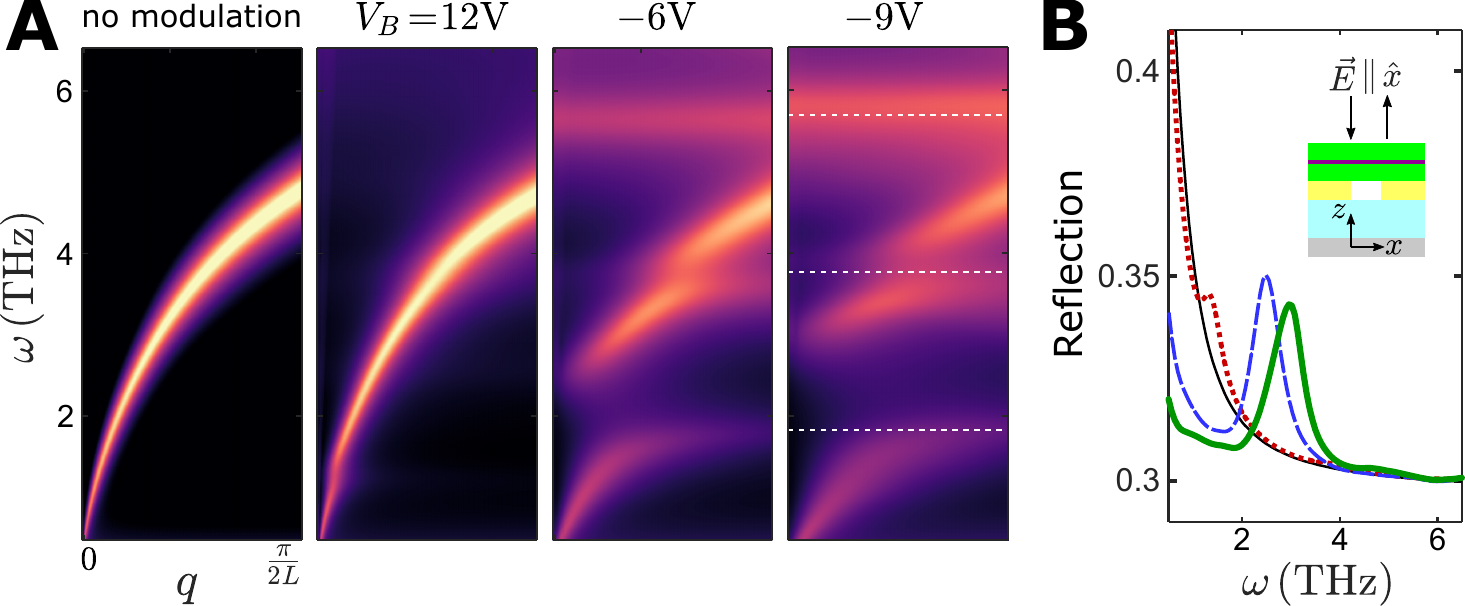}
    \caption{HIPP dispersion with ultra-strong coupling and far-field detection of HIPPs. \textbf{a} Density of states or $-\text{Im}\left[\text{Tr}\left([\epsilon(\mathbf{q},\omega)]^{-1}\right)\right]$ for visualizing the HIPP dispersion. The white dashed lines in $V_B=-9$V panel are denoting the same frequencies of the ISBT resonances ($\Delta j=1,2$, and 3) shown in Fig. \ref{fig.4.}\textbf{a}. \textbf{b} Reflection spectra for the normal incidence of light polarized along $x$-axis; solid thin black: no modulation ($E_F=0.15$eV), dotted red: $V_B=12$V, dashed blue: $-6$V, and solid thick green: $-9$V.}
    \label{fig.5.}
\end{figure}

Figure \ref{fig.5.}\textbf{a} shows the resulting HIPP dispersions featuring the ultra-strong coupling between the ISBT and the underlying plasmon-polaritons. If the system is spatially homogeneous, the polariton eigenmodes in graphene appear as the zeros of the scalar dynamical dielectric function $\epsilon(q,\omega)=1-q^2\sigma(q,\omega)/i\omega C(q,\omega)$  \cite{RPA, chinn}. Here, $C(q,\omega)$ is the dynamical capacitance of the system \cite{PCGP}, which connects the dynamic carrier density oscillation $\delta n(q,\omega)$ and the dynamic electric potential field on graphene $\delta U_E(q,\omega)$: $\delta U_E=\frac{e^2}{C}\delta n$. In essence, the dynamic capacitance $C$ encodes the information about the dielectric environment around graphene, in contrast to the conductivity $\sigma$ that encodes the dielectric property of graphene itself. With a periodic modulation along $x$-axis like in our system, the dynamical dielectric function is given as a matrix form  
\begin{equation}
\label{eq4}
[\epsilon(\mathbf{q},\omega)]_{m,m^\prime}=\delta_{m,m^\prime}- \sum_{l}C^{-1}(\mathbf{q}_m,\mathbf{q}_l;\omega)\frac{\mathbf{q}_{l}\cdot\mathbf{q}_{m^{\prime}}}{i\omega}\sigma_{xx}(\mathbf{q}_{l},\mathbf{q}_{m^{\prime}};\omega),
\end{equation}
where $m,m^\prime,l\in\mathbb{Z}$ are integer indices, $\mathbf{q}_m = \mathbf{q}+m G_0\hat{x}$ is the harmonic overtone of the polariton Bloch wavevector $\mathbf{q}$, and $C^{-1}$ is the inverse dynamic capacitance that governs a linear relation $\delta U_E(\mathbf{q}) = e^2\sum_{\mathbf{q}^\prime} C^{-1}(\mathbf{q},\mathbf{q}^\prime) \delta n(\mathbf{q}^\prime)$ (for more detailed explanation, see the Supplementary Information). Then, the polariton Bloch eigenmodes appear as the zeros of the determinant of the dynamical dielectric function matrix \cite{chinn,Minwoo_PRL,GP_mod3,PCGP}---i.e. the matrix $[\epsilon(\mathbf{q},\omega)]$ becomes non-invertible. Therefore, in Fig. \ref{fig.5.}\textbf{a}, we plotted the density of states approximated as $DOS(\mathbf{q},\omega)=-\text{Im}\left[\text{Tr}\left([\epsilon(\mathbf{q},\omega)]^{-1}\right)\right]$ to visualize the HIPP dispersion for polartions propagating along $x$-axis ($\mathbf{q}=q\hat{\mathbf{x}}$).

With a moderate depth of the SL modulation ($V_B=12$V case), the HIPP dispersion is similar to the plasmon-polariton dispersion with no modulation, and the ISBT feature is very subtle. As the SL modulation gets deeper, there emerge several HIPP branches resulting from the hybridization between the underlying plasmon-polartions and the ISBT resonances. With an extreme modulation ($V_B=-9$V case), we observe a huge Rabi-splitting ($\sim 2$THz) between the lowest branch and the second lowest, which is even comparable to the ISBT frequency itself ($\sim 2$THz). In such an ultra-strong coupling regime, a recent study reported that the electronic band structure of the material could be modified in return due to the vacuum fluctuation of the strongly-interacting polaritonic modes \cite{Ido2020}. We believe that our system would exhibit a similar behavior, but we didn't consider such additional corrections in this work. Figure \ref{fig.5.}\textbf{b} illustrates that this HIPP phenomenon can be detected in the far-field reflection as well. Each peak in the reflection spectra corresponds to the $q=0$ mode along the second lowest HIPP branch. In the reflection calculation, we assumed that the backgate substrate is silicon doped with a carrier density of $10^{15}$cm$^{-3}$. The diverging behavior at $\omega\rightarrow0$ for no-modulation and $V_B=12$V cases is due to the Drude response of the silicon backgate.

The emergent HIPPs found in our proposed system have several unique features compared to the usual intersubband-polaritons or intersubband-plasmon-polaritons studied in other platforms. First, the quantum well structure is given along the direction of the polariton propagation. Accordingly, the ISBT of our system occurs through in-plane electric fields along $x$-axis, which allows the far-field detection even with normal incidence of light. In contrast, conventional ISBT structures are based on vertical engineering of quantum wells \cite{Ido2020, IPP1, IPP2}, and the optical coupling requires out-of-plane electric fields. Second, as discussed earlier, the linear dispersion of Dirac electrons allows multiple (6$\sim$7) bands with equi-spaced energy levels to resonantly build up the ISBT strength. This resonant enhancement from multiple ladder-like bands below the Fermi surface is still possible with a quadratic dispersion of massive particles, but it will require a harmonic potential instead of a square potential. However, in realistic material platforms, it would remain as an extremely challenging task to engineer the quadratic shape of the harmonic potential precisely enough to maintain the uniform energy level spacing upto the 6th or 7th energy levels. Third, both the ISBT quanutm well structures and the plasmon-polaritons that couple to the ISBT are hosted simultaneously by monolayer graphene. This also contributes to the ultra-strong coupling, since the plasmon-polariton field strengths are by nature maximum at the plane of graphene. Similarly, in conventional vertically confined quantum well structures, when the ground state subband is populated, the 2D electron gas is naturally formed, hosting plasmon-polaritons confined around it \cite{IPP1}. But, both the ISBT ground state wavefunction and the plasmon-polariton fields have finite widths along $z$-axis, unlike our system where the subband states are confined at an atomically thin layer.

We emphasize again that the HIPPs shown in Fig. \ref{fig.5.} operate in the ultra-strongly coupled regime, featuring a giant Rabi-splitting that becomes comparable to the ISBT frequency. In this regime, several quantum electrodynamic phenomena can arise, such as material bandgap renormalization \cite{Ido2020} or anti-resonant coupling that breaks the rotating wave approximation and the Kubo conductivity formula \cite{ultrastrong}. Therefore, more precise determination of the HIPP dispersion would require a full quantum description of the ultra-strong coupling physics, which we leave as a future work. Lastly, even apart from the HIPP physics, the TIREDD-induced ladder-like energy bands themselves can be useful for high harmonic generations. Nonlinear optical responses can be resonantly enhanced by engineering the equi-spaced energy level of subbands in quantum well structures \cite{Lee2014}. As mentioned eariler, the linear dispersion of Dirac electrons naturally ensures the equi-spaced energy levels of the bound states in the square potential well. Thus, the 1D SL potential in graphene can be also used as a novel material platform for nonlinear optics.

In conclusion, our study suggests that the SL engineering in 2D materials can lead to the discovery of novel polariton phenomena emerging from the deformed electronic band structures. The modified subband structure of Dirac electrons under a 1D SL adds a completely new dimension to the polartion composition, leading to the formation of the HIPPs. This emergent HIPP is easily tunable by the double-gating scheme, provides a way to detect the SL-induced band structure changes with a far-field optical measurement, and becomes suitable for the study of quantum and nonlinear optics based on ultra-strong light-matter interaction.  Introducing a 2D SL \cite{nat_nano_SiO2} or patterning other 2D materials beyond graphene \cite{SL_TMD} could lead to more opportunities to study emergent polaritons with other novel formation mechanisms.

\begin{acknowledgments}
This work was supported by the Army Research Office (ARO) under a Grant No. W911NF-16-1-0319, and
by the National Science Foundation (NSF) under the Grants No. DMR-1741788 and DMR-1719875. M.J. was also supported in part by Cornell Fellowship and in part by the Kwanjeong Fellowship from Kwanjeong Educational Foundation.
\end{acknowledgments}


\begin{thebibliography}{35}%
\makeatletter
\providecommand \@ifxundefined [1]{%
 \@ifx{#1\undefined}
}%
\providecommand \@ifnum [1]{%
 \ifnum #1\expandafter \@firstoftwo
 \else \expandafter \@secondoftwo
 \fi
}%
\providecommand \@ifx [1]{%
 \ifx #1\expandafter \@firstoftwo
 \else \expandafter \@secondoftwo
 \fi
}%
\providecommand \natexlab [1]{#1}%
\providecommand \enquote  [1]{``#1''}%
\providecommand \bibnamefont  [1]{#1}%
\providecommand \bibfnamefont [1]{#1}%
\providecommand \citenamefont [1]{#1}%
\providecommand \href@noop [0]{\@secondoftwo}%
\providecommand \href [0]{\begingroup \@sanitize@url \@href}%
\providecommand \@href[1]{\@@startlink{#1}\@@href}%
\providecommand \@@href[1]{\endgroup#1\@@endlink}%
\providecommand \@sanitize@url [0]{\catcode `\\12\catcode `\$12\catcode
  `\&12\catcode `\#12\catcode `\^12\catcode `\_12\catcode `\%12\relax}%
\providecommand \@@startlink[1]{}%
\providecommand \@@endlink[0]{}%
\providecommand \url  [0]{\begingroup\@sanitize@url \@url }%
\providecommand \@url [1]{\endgroup\@href {#1}{\urlprefix }}%
\providecommand \urlprefix  [0]{URL }%
\providecommand \Eprint [0]{\href }%
\providecommand \doibase [0]{http://dx.doi.org/}%
\providecommand \selectlanguage [0]{\@gobble}%
\providecommand \bibinfo  [0]{\@secondoftwo}%
\providecommand \bibfield  [0]{\@secondoftwo}%
\providecommand \translation [1]{[#1]}%
\providecommand \BibitemOpen [0]{}%
\providecommand \bibitemStop [0]{}%
\providecommand \bibitemNoStop [0]{.\EOS\space}%
\providecommand \EOS [0]{\spacefactor3000\relax}%
\providecommand \BibitemShut  [1]{\csname bibitem#1\endcsname}%
\let\auto@bib@innerbib\@empty
\bibitem [{\citenamefont {Park}\ \emph {et~al.}(2008)\citenamefont {Park},
  \citenamefont {Yang}, \citenamefont {Son}, \citenamefont {Cohen},\ and\
  \citenamefont {Louie}}]{BandEngineer1}%
  \BibitemOpen
  \bibfield  {author} {\bibinfo {author} {\bibfnamefont {C.-H.}\ \bibnamefont
  {Park}}, \bibinfo {author} {\bibfnamefont {L.}~\bibnamefont {Yang}}, \bibinfo
  {author} {\bibfnamefont {Y.-W.}\ \bibnamefont {Son}}, \bibinfo {author}
  {\bibfnamefont {M.~L.}\ \bibnamefont {Cohen}}, \ and\ \bibinfo {author}
  {\bibfnamefont {S.~G.}\ \bibnamefont {Louie}},\ }\href {\doibase
  10.1038/nphys890} {\bibfield  {journal} {\bibinfo  {journal} {Nat. Phys.}\
  }\textbf {\bibinfo {volume} {4}},\ \bibinfo {pages} {213–217} (\bibinfo
  {year} {2008})}\BibitemShut {NoStop}%
\bibitem [{\citenamefont {Park}\ \emph {et~al.}(2009)\citenamefont {Park},
  \citenamefont {Son}, \citenamefont {Yang}, \citenamefont {Cohen},\ and\
  \citenamefont {Louie}}]{BandEngineer2}%
  \BibitemOpen
  \bibfield  {author} {\bibinfo {author} {\bibfnamefont {C.-H.}\ \bibnamefont
  {Park}}, \bibinfo {author} {\bibfnamefont {Y.-W.}\ \bibnamefont {Son}},
  \bibinfo {author} {\bibfnamefont {L.}~\bibnamefont {Yang}}, \bibinfo {author}
  {\bibfnamefont {M.~L.}\ \bibnamefont {Cohen}}, \ and\ \bibinfo {author}
  {\bibfnamefont {S.~G.}\ \bibnamefont {Louie}},\ }\href {\doibase
  10.1103/PhysRevLett.103.046808} {\bibfield  {journal} {\bibinfo  {journal}
  {Phys. Rev. Lett.}\ }\textbf {\bibinfo {volume} {103}},\ \bibinfo {pages}
  {046808} (\bibinfo {year} {2009})}\BibitemShut {NoStop}%
\bibitem [{\citenamefont {Brey}\ and\ \citenamefont
  {Fertig}(2009)}]{BandEngineer3}%
  \BibitemOpen
  \bibfield  {author} {\bibinfo {author} {\bibfnamefont {L.}~\bibnamefont
  {Brey}}\ and\ \bibinfo {author} {\bibfnamefont {H.~A.}\ \bibnamefont
  {Fertig}},\ }\href {\doibase 10.1103/PhysRevLett.103.046809} {\bibfield
  {journal} {\bibinfo  {journal} {Phys. Rev. Lett.}\ }\textbf {\bibinfo
  {volume} {103}},\ \bibinfo {pages} {046809} (\bibinfo {year}
  {2009})}\BibitemShut {NoStop}%
\bibitem [{\citenamefont {Forsythe}\ \emph {et~al.}(2018)\citenamefont
  {Forsythe}, \citenamefont {Zhou}, \citenamefont {Watanabe}, \citenamefont
  {Taniguchi}, \citenamefont {Pasupathy}, \citenamefont {Moon}, \citenamefont
  {Koshino}, \citenamefont {Kim},\ and\ \citenamefont {Dean}}]{nat_nano_SiO2}%
  \BibitemOpen
  \bibfield  {author} {\bibinfo {author} {\bibfnamefont {C.}~\bibnamefont
  {Forsythe}}, \bibinfo {author} {\bibfnamefont {X.}~\bibnamefont {Zhou}},
  \bibinfo {author} {\bibfnamefont {K.}~\bibnamefont {Watanabe}}, \bibinfo
  {author} {\bibfnamefont {T.}~\bibnamefont {Taniguchi}}, \bibinfo {author}
  {\bibfnamefont {A.}~\bibnamefont {Pasupathy}}, \bibinfo {author}
  {\bibfnamefont {P.}~\bibnamefont {Moon}}, \bibinfo {author} {\bibfnamefont
  {M.}~\bibnamefont {Koshino}}, \bibinfo {author} {\bibfnamefont
  {P.}~\bibnamefont {Kim}}, \ and\ \bibinfo {author} {\bibfnamefont {C.~R.}\
  \bibnamefont {Dean}},\ }\href {\doibase 10.1038/s41565-018-0138-7} {\bibfield
   {journal} {\bibinfo  {journal} {Nat. Nanotechnol.}\ }\textbf {\bibinfo
  {volume} {13}},\ \bibinfo {pages} {566–571} (\bibinfo {year}
  {2018})}\BibitemShut {NoStop}%
\bibitem [{\citenamefont {Li}\ \emph {et~al.}(2021)\citenamefont {Li},
  \citenamefont {Dietrich}, \citenamefont {Forsythe}, \citenamefont
  {Taniguchi}, \citenamefont {Watanabe}, \citenamefont {Moon},\ and\
  \citenamefont {Dean}}]{BandEngineer5}%
  \BibitemOpen
  \bibfield  {author} {\bibinfo {author} {\bibfnamefont {Y.}~\bibnamefont
  {Li}}, \bibinfo {author} {\bibfnamefont {S.}~\bibnamefont {Dietrich}},
  \bibinfo {author} {\bibfnamefont {C.}~\bibnamefont {Forsythe}}, \bibinfo
  {author} {\bibfnamefont {T.}~\bibnamefont {Taniguchi}}, \bibinfo {author}
  {\bibfnamefont {K.}~\bibnamefont {Watanabe}}, \bibinfo {author}
  {\bibfnamefont {P.}~\bibnamefont {Moon}}, \ and\ \bibinfo {author}
  {\bibfnamefont {C.~R.}\ \bibnamefont {Dean}},\ }\href {\doibase
  10.1038/s41565-021-00849-9} {\bibfield  {journal} {\bibinfo  {journal} {Nat.
  Nanotechnol.}\ }\textbf {\bibinfo {volume} {16}},\ \bibinfo {pages}
  {525–530} (\bibinfo {year} {2021})}\BibitemShut {NoStop}%
\bibitem [{\citenamefont {Brey}\ \emph {et~al.}(2020)\citenamefont {Brey},
  \citenamefont {Stauber}, \citenamefont {Martín-Moreno},\ and\ \citenamefont
  {Gómez-Santos}}]{Brey_PRL}%
  \BibitemOpen
  \bibfield  {author} {\bibinfo {author} {\bibfnamefont {L.}~\bibnamefont
  {Brey}}, \bibinfo {author} {\bibfnamefont {T.}~\bibnamefont {Stauber}},
  \bibinfo {author} {\bibfnamefont {L.}~\bibnamefont {Martín-Moreno}}, \ and\
  \bibinfo {author} {\bibfnamefont {G.}~\bibnamefont {Gómez-Santos}},\ }\href
  {\doibase 10.1103/PhysRevLett.124.257401} {\bibfield  {journal} {\bibinfo
  {journal} {Phys. Rev. Lett.}\ }\textbf {\bibinfo {volume} {124}},\ \bibinfo
  {pages} {257401} (\bibinfo {year} {2020})}\BibitemShut {NoStop}%
\bibitem [{\citenamefont {Woessner}\ \emph {et~al.}(2015)\citenamefont
  {Woessner}, \citenamefont {Lundeberg}, \citenamefont {Gao}, \citenamefont
  {Principi}, \citenamefont {Alonso-Gonzalez}, \citenamefont {Carrega},
  \citenamefont {Watanabe}, \citenamefont {Taniguchi}, \citenamefont {Vignale},
  \citenamefont {Polini}, \citenamefont {Hone}, \citenamefont {Hillenbrand},\
  and\ \citenamefont {Koppens}}]{hBN1}%
  \BibitemOpen
  \bibfield  {author} {\bibinfo {author} {\bibfnamefont {A.}~\bibnamefont
  {Woessner}}, \bibinfo {author} {\bibfnamefont {M.~B.}\ \bibnamefont
  {Lundeberg}}, \bibinfo {author} {\bibfnamefont {Y.}~\bibnamefont {Gao}},
  \bibinfo {author} {\bibfnamefont {A.}~\bibnamefont {Principi}}, \bibinfo
  {author} {\bibfnamefont {P.}~\bibnamefont {Alonso-Gonzalez}}, \bibinfo
  {author} {\bibfnamefont {M.}~\bibnamefont {Carrega}}, \bibinfo {author}
  {\bibfnamefont {K.}~\bibnamefont {Watanabe}}, \bibinfo {author}
  {\bibfnamefont {T.}~\bibnamefont {Taniguchi}}, \bibinfo {author}
  {\bibfnamefont {G.}~\bibnamefont {Vignale}}, \bibinfo {author} {\bibfnamefont
  {M.}~\bibnamefont {Polini}}, \bibinfo {author} {\bibfnamefont
  {J.}~\bibnamefont {Hone}}, \bibinfo {author} {\bibfnamefont {R.}~\bibnamefont
  {Hillenbrand}}, \ and\ \bibinfo {author} {\bibfnamefont {F.~H.~L.}\
  \bibnamefont {Koppens}},\ }\href {\doibase 10.1038/nmat4169} {\bibfield
  {journal} {\bibinfo  {journal} {Nature Mater.}\ }\textbf {\bibinfo {volume}
  {14}},\ \bibinfo {pages} {421} (\bibinfo {year} {2015})}\BibitemShut
  {NoStop}%
\bibitem [{\citenamefont {Alonso-Gonzalez}\ \emph {et~al.}(2017)\citenamefont
  {Alonso-Gonzalez}, \citenamefont {Nikitin}, \citenamefont {Gao},
  \citenamefont {Woessner}, \citenamefont {Lundeberg}, \citenamefont
  {Principi}, \citenamefont {Forcellini}, \citenamefont {Yan}, \citenamefont
  {Velez}, \citenamefont {Huber}, \citenamefont {Watanabe}, \citenamefont
  {Taniguchi}, \citenamefont {Casanova}, \citenamefont {Hueso}, \citenamefont
  {Polini}, \citenamefont {Hone}, \citenamefont {Koppens},\ and\ \citenamefont
  {Hillenbrand}}]{hBN3}%
  \BibitemOpen
  \bibfield  {author} {\bibinfo {author} {\bibfnamefont {P.}~\bibnamefont
  {Alonso-Gonzalez}}, \bibinfo {author} {\bibfnamefont {A.~Y.}\ \bibnamefont
  {Nikitin}}, \bibinfo {author} {\bibfnamefont {Y.}~\bibnamefont {Gao}},
  \bibinfo {author} {\bibfnamefont {A.}~\bibnamefont {Woessner}}, \bibinfo
  {author} {\bibfnamefont {M.~B.}\ \bibnamefont {Lundeberg}}, \bibinfo {author}
  {\bibfnamefont {A.}~\bibnamefont {Principi}}, \bibinfo {author}
  {\bibfnamefont {N.}~\bibnamefont {Forcellini}}, \bibinfo {author}
  {\bibfnamefont {W.}~\bibnamefont {Yan}}, \bibinfo {author} {\bibfnamefont
  {S.}~\bibnamefont {Velez}}, \bibinfo {author} {\bibfnamefont {A.~J.}\
  \bibnamefont {Huber}}, \bibinfo {author} {\bibfnamefont {K.}~\bibnamefont
  {Watanabe}}, \bibinfo {author} {\bibfnamefont {T.}~\bibnamefont {Taniguchi}},
  \bibinfo {author} {\bibfnamefont {F.}~\bibnamefont {Casanova}}, \bibinfo
  {author} {\bibfnamefont {L.~E.}\ \bibnamefont {Hueso}}, \bibinfo {author}
  {\bibfnamefont {M.}~\bibnamefont {Polini}}, \bibinfo {author} {\bibfnamefont
  {J.}~\bibnamefont {Hone}}, \bibinfo {author} {\bibfnamefont {F.~H.~L.}\
  \bibnamefont {Koppens}}, \ and\ \bibinfo {author} {\bibfnamefont
  {R.}~\bibnamefont {Hillenbrand}},\ }\href {\doibase 10.1038/nnano.2016.185}
  {\bibfield  {journal} {\bibinfo  {journal} {Nat. Nanotech.}\ }\textbf
  {\bibinfo {volume} {12}},\ \bibinfo {pages} {31} (\bibinfo {year}
  {2017})}\BibitemShut {NoStop}%
\bibitem [{\citenamefont {Sidler}\ \emph {et~al.}(2017)\citenamefont {Sidler},
  \citenamefont {Back}, \citenamefont {Cotlet}, \citenamefont {Srivastava},
  \citenamefont {Fink}, \citenamefont {Kroner}, \citenamefont {Demler},\ and\
  \citenamefont {Imamoglu}}]{ETP1}%
  \BibitemOpen
  \bibfield  {author} {\bibinfo {author} {\bibfnamefont {M.}~\bibnamefont
  {Sidler}}, \bibinfo {author} {\bibfnamefont {P.}~\bibnamefont {Back}},
  \bibinfo {author} {\bibfnamefont {O.}~\bibnamefont {Cotlet}}, \bibinfo
  {author} {\bibfnamefont {A.}~\bibnamefont {Srivastava}}, \bibinfo {author}
  {\bibfnamefont {T.}~\bibnamefont {Fink}}, \bibinfo {author} {\bibfnamefont
  {M.}~\bibnamefont {Kroner}}, \bibinfo {author} {\bibfnamefont
  {E.}~\bibnamefont {Demler}}, \ and\ \bibinfo {author} {\bibfnamefont
  {A.}~\bibnamefont {Imamoglu}},\ }\href {\doibase 10.1038/nphys3949}
  {\bibfield  {journal} {\bibinfo  {journal} {Nat. Phys.}\ }\textbf {\bibinfo
  {volume} {13}},\ \bibinfo {pages} {255} (\bibinfo {year} {2017})}\BibitemShut
  {NoStop}%
\bibitem [{\citenamefont {Koksal}\ \emph {et~al.}(2021)\citenamefont {Koksal},
  \citenamefont {Jung}, \citenamefont {Manolatou}, \citenamefont {Vamivakas},
  \citenamefont {Shvets},\ and\ \citenamefont {Rana}}]{ETP2}%
  \BibitemOpen
  \bibfield  {author} {\bibinfo {author} {\bibfnamefont {O.}~\bibnamefont
  {Koksal}}, \bibinfo {author} {\bibfnamefont {M.}~\bibnamefont {Jung}},
  \bibinfo {author} {\bibfnamefont {C.}~\bibnamefont {Manolatou}}, \bibinfo
  {author} {\bibfnamefont {A.~N.}\ \bibnamefont {Vamivakas}}, \bibinfo {author}
  {\bibfnamefont {G.}~\bibnamefont {Shvets}}, \ and\ \bibinfo {author}
  {\bibfnamefont {F.}~\bibnamefont {Rana}},\ }\href {\doibase
  10.1103/PhysRevResearch.3.033064} {\bibfield  {journal} {\bibinfo  {journal}
  {Phys. Rev. Res.}\ }\textbf {\bibinfo {volume} {3}},\ \bibinfo {pages}
  {033064} (\bibinfo {year} {2021})}\BibitemShut {NoStop}%
\bibitem [{\citenamefont {Ni}\ \emph {et~al.}(2016)\citenamefont {Ni},
  \citenamefont {Wang}, \citenamefont {Goldflam}, \citenamefont {Wagner},
  \citenamefont {Fei}, \citenamefont {McLeod}, \citenamefont {Liu},
  \citenamefont {Keilmann}, \citenamefont {Özyilmaz}, \citenamefont {Neto},
  \citenamefont {Hone}, \citenamefont {Fogler},\ and\ \citenamefont
  {Basov}}]{photo_mod1}%
  \BibitemOpen
  \bibfield  {author} {\bibinfo {author} {\bibfnamefont {G.~X.}\ \bibnamefont
  {Ni}}, \bibinfo {author} {\bibfnamefont {L.}~\bibnamefont {Wang}}, \bibinfo
  {author} {\bibfnamefont {M.~D.}\ \bibnamefont {Goldflam}}, \bibinfo {author}
  {\bibfnamefont {M.}~\bibnamefont {Wagner}}, \bibinfo {author} {\bibfnamefont
  {Z.}~\bibnamefont {Fei}}, \bibinfo {author} {\bibfnamefont {A.~S.}\
  \bibnamefont {McLeod}}, \bibinfo {author} {\bibfnamefont {M.~K.}\
  \bibnamefont {Liu}}, \bibinfo {author} {\bibfnamefont {F.}~\bibnamefont
  {Keilmann}}, \bibinfo {author} {\bibfnamefont {B.}~\bibnamefont {Özyilmaz}},
  \bibinfo {author} {\bibfnamefont {A.~H.~C.}\ \bibnamefont {Neto}}, \bibinfo
  {author} {\bibfnamefont {J.}~\bibnamefont {Hone}}, \bibinfo {author}
  {\bibfnamefont {M.~M.}\ \bibnamefont {Fogler}}, \ and\ \bibinfo {author}
  {\bibfnamefont {D.~N.}\ \bibnamefont {Basov}},\ }\href {\doibase
  10.1038/nphoton.2016.45} {\bibfield  {journal} {\bibinfo  {journal} {Nat.
  Photon.}\ }\textbf {\bibinfo {volume} {10}},\ \bibinfo {pages} {244}
  (\bibinfo {year} {2016})}\BibitemShut {NoStop}%
\bibitem [{\citenamefont {Tan}\ \emph {et~al.}(2020)\citenamefont {Tan},
  \citenamefont {Cotlet}, \citenamefont {Bergschneider}, \citenamefont
  {Schmidt}, \citenamefont {Back}, \citenamefont {Shimazaki}, \citenamefont
  {Kroner},\ and\ \citenamefont {İmamoğlu}}]{photo_mod2}%
  \BibitemOpen
  \bibfield  {author} {\bibinfo {author} {\bibfnamefont {L.~B.}\ \bibnamefont
  {Tan}}, \bibinfo {author} {\bibfnamefont {O.}~\bibnamefont {Cotlet}},
  \bibinfo {author} {\bibfnamefont {A.}~\bibnamefont {Bergschneider}}, \bibinfo
  {author} {\bibfnamefont {R.}~\bibnamefont {Schmidt}}, \bibinfo {author}
  {\bibfnamefont {P.}~\bibnamefont {Back}}, \bibinfo {author} {\bibfnamefont
  {Y.}~\bibnamefont {Shimazaki}}, \bibinfo {author} {\bibfnamefont
  {M.}~\bibnamefont {Kroner}}, \ and\ \bibinfo {author} {\bibfnamefont
  {A.}~\bibnamefont {İmamoğlu}},\ }\href {\doibase
  10.1103/PhysRevX.10.021011} {\bibfield  {journal} {\bibinfo  {journal} {Phys.
  Rev. X}\ }\textbf {\bibinfo {volume} {10}},\ \bibinfo {pages} {021011}
  (\bibinfo {year} {2020})}\BibitemShut {NoStop}%
\bibitem [{\citenamefont {Sternbach}\ \emph {et~al.}(2020)\citenamefont
  {Sternbach}, \citenamefont {Chae}, \citenamefont {Latini}, \citenamefont
  {Rikhter}, \citenamefont {Shao}, \citenamefont {Li}, \citenamefont {Rhodes},
  \citenamefont {Kim}, \citenamefont {Schuck}, \citenamefont {Xu},
  \citenamefont {Zhu}, \citenamefont {Averitt}, \citenamefont {Hone},
  \citenamefont {Fogler}, \citenamefont {Rubio},\ and\ \citenamefont
  {Basov}}]{photo_mod3}%
  \BibitemOpen
  \bibfield  {author} {\bibinfo {author} {\bibfnamefont {A.~J.}\ \bibnamefont
  {Sternbach}}, \bibinfo {author} {\bibfnamefont {S.~H.}\ \bibnamefont {Chae}},
  \bibinfo {author} {\bibfnamefont {S.}~\bibnamefont {Latini}}, \bibinfo
  {author} {\bibfnamefont {A.~A.}\ \bibnamefont {Rikhter}}, \bibinfo {author}
  {\bibfnamefont {Y.}~\bibnamefont {Shao}}, \bibinfo {author} {\bibfnamefont
  {B.}~\bibnamefont {Li}}, \bibinfo {author} {\bibfnamefont {D.}~\bibnamefont
  {Rhodes}}, \bibinfo {author} {\bibfnamefont {B.}~\bibnamefont {Kim}},
  \bibinfo {author} {\bibfnamefont {P.~J.}\ \bibnamefont {Schuck}}, \bibinfo
  {author} {\bibfnamefont {X.}~\bibnamefont {Xu}}, \bibinfo {author}
  {\bibfnamefont {X.-Y.}\ \bibnamefont {Zhu}}, \bibinfo {author} {\bibfnamefont
  {R.~D.}\ \bibnamefont {Averitt}}, \bibinfo {author} {\bibfnamefont
  {J.}~\bibnamefont {Hone}}, \bibinfo {author} {\bibfnamefont {M.~M.}\
  \bibnamefont {Fogler}}, \bibinfo {author} {\bibfnamefont {A.}~\bibnamefont
  {Rubio}}, \ and\ \bibinfo {author} {\bibfnamefont {D.~N.}\ \bibnamefont
  {Basov}},\ }\href {\doibase 10.1126/science.abe9163} {\bibfield  {journal}
  {\bibinfo  {journal} {Science}\ }\textbf {\bibinfo {volume} {371}},\ \bibinfo
  {pages} {617} (\bibinfo {year} {2020})}\BibitemShut {NoStop}%
\bibitem [{\citenamefont {Rana}\ \emph {et~al.}(2021)\citenamefont {Rana},
  \citenamefont {Koksal}, \citenamefont {Jung}, \citenamefont {Shvets},
  \citenamefont {Vamivakas},\ and\ \citenamefont {Manolatou}}]{ETP_mod}%
  \BibitemOpen
  \bibfield  {author} {\bibinfo {author} {\bibfnamefont {F.}~\bibnamefont
  {Rana}}, \bibinfo {author} {\bibfnamefont {O.}~\bibnamefont {Koksal}},
  \bibinfo {author} {\bibfnamefont {M.}~\bibnamefont {Jung}}, \bibinfo {author}
  {\bibfnamefont {G.}~\bibnamefont {Shvets}}, \bibinfo {author} {\bibfnamefont
  {A.~N.}\ \bibnamefont {Vamivakas}}, \ and\ \bibinfo {author} {\bibfnamefont
  {C.}~\bibnamefont {Manolatou}},\ }\href {\doibase
  10.1103/PhysRevLett.126.127402} {\bibfield  {journal} {\bibinfo  {journal}
  {Phys. Rev. Lett.}\ }\textbf {\bibinfo {volume} {126}},\ \bibinfo {pages}
  {127402} (\bibinfo {year} {2021})}\BibitemShut {NoStop}%
\bibitem [{\citenamefont {Woessner}\ \emph {et~al.}(2017)\citenamefont
  {Woessner}, \citenamefont {Gao}, \citenamefont {Torre}, \citenamefont
  {Lundeberg}, \citenamefont {Tan}, \citenamefont {Watanabe}, \citenamefont
  {Taniguchi}, \citenamefont {Hillenbrand}, \citenamefont {Hone}, \citenamefont
  {Polini},\ and\ \citenamefont {Koppens}}]{hBN4}%
  \BibitemOpen
  \bibfield  {author} {\bibinfo {author} {\bibfnamefont {A.}~\bibnamefont
  {Woessner}}, \bibinfo {author} {\bibfnamefont {Y.}~\bibnamefont {Gao}},
  \bibinfo {author} {\bibfnamefont {I.}~\bibnamefont {Torre}}, \bibinfo
  {author} {\bibfnamefont {M.~B.}\ \bibnamefont {Lundeberg}}, \bibinfo {author}
  {\bibfnamefont {C.}~\bibnamefont {Tan}}, \bibinfo {author} {\bibfnamefont
  {K.}~\bibnamefont {Watanabe}}, \bibinfo {author} {\bibfnamefont
  {T.}~\bibnamefont {Taniguchi}}, \bibinfo {author} {\bibfnamefont
  {R.}~\bibnamefont {Hillenbrand}}, \bibinfo {author} {\bibfnamefont
  {J.}~\bibnamefont {Hone}}, \bibinfo {author} {\bibfnamefont {M.}~\bibnamefont
  {Polini}}, \ and\ \bibinfo {author} {\bibfnamefont {F.~H.~L.}\ \bibnamefont
  {Koppens}},\ }\href {\doibase 10.1038/nphoton.2017.98} {\bibfield  {journal}
  {\bibinfo  {journal} {Nat. Photonics}\ }\textbf {\bibinfo {volume} {11}},\
  \bibinfo {pages} {421} (\bibinfo {year} {2017})}\BibitemShut {NoStop}%
\bibitem [{\citenamefont {Fan}\ \emph {et~al.}(2019)\citenamefont {Fan},
  \citenamefont {Dutta-Gupta}, \citenamefont {Gladstone}, \citenamefont
  {Trendafilov}, \citenamefont {Bosch}, \citenamefont {Jung}, \citenamefont
  {Iyer}, \citenamefont {Giles}, \citenamefont {Shcherbakov}, \citenamefont
  {Feigelson}, \citenamefont {Caldwell}, \citenamefont {Allen}, \citenamefont
  {Allen},\ and\ \citenamefont {Shvets}}]{GP_mod2}%
  \BibitemOpen
  \bibfield  {author} {\bibinfo {author} {\bibfnamefont {Z.}~\bibnamefont
  {Fan}}, \bibinfo {author} {\bibfnamefont {S.}~\bibnamefont {Dutta-Gupta}},
  \bibinfo {author} {\bibfnamefont {R.}~\bibnamefont {Gladstone}}, \bibinfo
  {author} {\bibfnamefont {S.}~\bibnamefont {Trendafilov}}, \bibinfo {author}
  {\bibfnamefont {M.}~\bibnamefont {Bosch}}, \bibinfo {author} {\bibfnamefont
  {M.}~\bibnamefont {Jung}}, \bibinfo {author} {\bibfnamefont {G.~R.~S.}\
  \bibnamefont {Iyer}}, \bibinfo {author} {\bibfnamefont {A.~J.}\ \bibnamefont
  {Giles}}, \bibinfo {author} {\bibfnamefont {M.}~\bibnamefont {Shcherbakov}},
  \bibinfo {author} {\bibfnamefont {B.}~\bibnamefont {Feigelson}}, \bibinfo
  {author} {\bibfnamefont {J.~D.}\ \bibnamefont {Caldwell}}, \bibinfo {author}
  {\bibfnamefont {M.}~\bibnamefont {Allen}}, \bibinfo {author} {\bibfnamefont
  {J.}~\bibnamefont {Allen}}, \ and\ \bibinfo {author} {\bibfnamefont
  {G.}~\bibnamefont {Shvets}},\ }\href {\doibase 10.1515/nanoph-2019-0108}
  {\bibfield  {journal} {\bibinfo  {journal} {Nanophotonics}\ }\textbf
  {\bibinfo {volume} {8}},\ \bibinfo {pages} {1417} (\bibinfo {year}
  {2019})}\BibitemShut {NoStop}%
\bibitem [{\citenamefont {Xiong}\ \emph {et~al.}(2021)\citenamefont {Xiong},
  \citenamefont {Li}, \citenamefont {Jung}, \citenamefont {Forsythe},
  \citenamefont {Zhang}, \citenamefont {McLeod}, \citenamefont {Dong},
  \citenamefont {Liu}, \citenamefont {Ruta}, \citenamefont {Li}, \citenamefont
  {Watanabe}, \citenamefont {Taniguchi}, \citenamefont {Fogler}, \citenamefont
  {Edgar}, \citenamefont {Shvets},\ and\ \citenamefont {adn
  DN~Basov}}]{GP_mod3}%
  \BibitemOpen
  \bibfield  {author} {\bibinfo {author} {\bibfnamefont {L.}~\bibnamefont
  {Xiong}}, \bibinfo {author} {\bibfnamefont {Y.}~\bibnamefont {Li}}, \bibinfo
  {author} {\bibfnamefont {M.}~\bibnamefont {Jung}}, \bibinfo {author}
  {\bibfnamefont {C.}~\bibnamefont {Forsythe}}, \bibinfo {author}
  {\bibfnamefont {S.}~\bibnamefont {Zhang}}, \bibinfo {author} {\bibfnamefont
  {A.~S.}\ \bibnamefont {McLeod}}, \bibinfo {author} {\bibfnamefont
  {Y.}~\bibnamefont {Dong}}, \bibinfo {author} {\bibfnamefont {S.}~\bibnamefont
  {Liu}}, \bibinfo {author} {\bibfnamefont {F.~L.}\ \bibnamefont {Ruta}},
  \bibinfo {author} {\bibfnamefont {C.}~\bibnamefont {Li}}, \bibinfo {author}
  {\bibfnamefont {K.}~\bibnamefont {Watanabe}}, \bibinfo {author}
  {\bibfnamefont {T.}~\bibnamefont {Taniguchi}}, \bibinfo {author}
  {\bibfnamefont {M.~M.}\ \bibnamefont {Fogler}}, \bibinfo {author}
  {\bibfnamefont {J.~H.}\ \bibnamefont {Edgar}}, \bibinfo {author}
  {\bibfnamefont {G.}~\bibnamefont {Shvets}}, \ and\ \bibinfo {author}
  {\bibfnamefont {C.~R.~D.}\ \bibnamefont {adn DN~Basov}},\ }\href {\doibase
  10.1126/sciadv.abe8087} {\bibfield  {journal} {\bibinfo  {journal} {Sci.
  Adv.}\ }\textbf {\bibinfo {volume} {7}},\ \bibinfo {pages} {eabe8087}
  (\bibinfo {year} {2021})}\BibitemShut {NoStop}%
\bibitem [{\citenamefont {Fang}\ \emph {et~al.}(2007)\citenamefont {Fang},
  \citenamefont {Konar}, \citenamefont {Xing},\ and\ \citenamefont
  {Jena}}]{QCG}%
  \BibitemOpen
  \bibfield  {author} {\bibinfo {author} {\bibfnamefont {T.}~\bibnamefont
  {Fang}}, \bibinfo {author} {\bibfnamefont {A.}~\bibnamefont {Konar}},
  \bibinfo {author} {\bibfnamefont {H.}~\bibnamefont {Xing}}, \ and\ \bibinfo
  {author} {\bibfnamefont {D.}~\bibnamefont {Jena}},\ }\href {\doibase
  10.1063/1.2776887} {\bibfield  {journal} {\bibinfo  {journal} {Appl. Phys.
  Lett.}\ }\textbf {\bibinfo {volume} {91}},\ \bibinfo {pages} {092109}
  (\bibinfo {year} {2007})}\BibitemShut {NoStop}%
\bibitem [{\citenamefont {Jung}\ \emph {et~al.}(2018)\citenamefont {Jung},
  \citenamefont {Fan},\ and\ \citenamefont {Shvets}}]{Minwoo_PRL}%
  \BibitemOpen
  \bibfield  {author} {\bibinfo {author} {\bibfnamefont {M.}~\bibnamefont
  {Jung}}, \bibinfo {author} {\bibfnamefont {Z.}~\bibnamefont {Fan}}, \ and\
  \bibinfo {author} {\bibfnamefont {G.}~\bibnamefont {Shvets}},\ }\href
  {\doibase 10.1103/PhysRevLett.121.086807} {\bibfield  {journal} {\bibinfo
  {journal} {Phys. Rev. Lett.}\ }\textbf {\bibinfo {volume} {121}},\ \bibinfo
  {pages} {086807} (\bibinfo {year} {2018})}\BibitemShut {NoStop}%
\bibitem [{\citenamefont {Yu}\ \emph {et~al.}(2013)\citenamefont {Yu},
  \citenamefont {Jalil}, \citenamefont {Belle}, \citenamefont {Mayorov},
  \citenamefont {Blake}, \citenamefont {Schedin}, \citenamefont {Morozov},
  \citenamefont {Ponomarenko}, \citenamefont {Chiappini}, \citenamefont
  {Wiedmann}, \citenamefont {Zeitler}, \citenamefont {Katsnelson},
  \citenamefont {Geim}, \citenamefont {Novoselov}, ,\ and\ \citenamefont
  {Elias}}]{novoselov_pnas13}%
  \BibitemOpen
  \bibfield  {author} {\bibinfo {author} {\bibfnamefont {G.~L.}\ \bibnamefont
  {Yu}}, \bibinfo {author} {\bibfnamefont {R.}~\bibnamefont {Jalil}}, \bibinfo
  {author} {\bibfnamefont {B.}~\bibnamefont {Belle}}, \bibinfo {author}
  {\bibfnamefont {A.~S.}\ \bibnamefont {Mayorov}}, \bibinfo {author}
  {\bibfnamefont {P.}~\bibnamefont {Blake}}, \bibinfo {author} {\bibfnamefont
  {F.}~\bibnamefont {Schedin}}, \bibinfo {author} {\bibfnamefont {S.~V.}\
  \bibnamefont {Morozov}}, \bibinfo {author} {\bibfnamefont {L.~A.}\
  \bibnamefont {Ponomarenko}}, \bibinfo {author} {\bibfnamefont
  {F.}~\bibnamefont {Chiappini}}, \bibinfo {author} {\bibfnamefont
  {S.}~\bibnamefont {Wiedmann}}, \bibinfo {author} {\bibfnamefont
  {U.}~\bibnamefont {Zeitler}}, \bibinfo {author} {\bibfnamefont {M.~I.}\
  \bibnamefont {Katsnelson}}, \bibinfo {author} {\bibfnamefont {A.~K.}\
  \bibnamefont {Geim}}, \bibinfo {author} {\bibfnamefont {K.~S.}\ \bibnamefont
  {Novoselov}}, , \ and\ \bibinfo {author} {\bibfnamefont {D.~C.}\ \bibnamefont
  {Elias}},\ }\href {\doibase 10.1073/pnas.1300599110} {\bibfield  {journal}
  {\bibinfo  {journal} {Proc. Natl. Acad. Sci. U.S.A.}\ }\textbf {\bibinfo
  {volume} {110}},\ \bibinfo {pages} {3282} (\bibinfo {year}
  {2013})}\BibitemShut {NoStop}%
\bibitem [{\citenamefont {Katsnelson}\ \emph {et~al.}(2006)\citenamefont
  {Katsnelson}, \citenamefont {Novoselov},\ and\ \citenamefont
  {Geim}}]{Klein1}%
  \BibitemOpen
  \bibfield  {author} {\bibinfo {author} {\bibfnamefont {M.~I.}\ \bibnamefont
  {Katsnelson}}, \bibinfo {author} {\bibfnamefont {K.~S.}\ \bibnamefont
  {Novoselov}}, \ and\ \bibinfo {author} {\bibfnamefont {A.~K.}\ \bibnamefont
  {Geim}},\ }\href {\doibase 10.1038/nphys384} {\bibfield  {journal} {\bibinfo
  {journal} {Nat. Phys.}\ }\textbf {\bibinfo {volume} {2}},\ \bibinfo {pages}
  {620} (\bibinfo {year} {2006})}\BibitemShut {NoStop}%
\bibitem [{\citenamefont {Young}\ and\ \citenamefont {Kim}(2009)}]{Klein2}%
  \BibitemOpen
  \bibfield  {author} {\bibinfo {author} {\bibfnamefont {A.~F.}\ \bibnamefont
  {Young}}\ and\ \bibinfo {author} {\bibfnamefont {P.}~\bibnamefont {Kim}},\
  }\href {\doibase 10.1038/nphys1198} {\bibfield  {journal} {\bibinfo
  {journal} {Nat. Phys.}\ }\textbf {\bibinfo {volume} {5}},\ \bibinfo {pages}
  {222} (\bibinfo {year} {2009})}\BibitemShut {NoStop}%
\bibitem [{\citenamefont {Wang}\ \emph {et~al.}(2013)\citenamefont {Wang},
  \citenamefont {Meric}, \citenamefont {Huang}, \citenamefont {Gao},
  \citenamefont {Gao}, \citenamefont {Tran}, \citenamefont {Taniguchi},
  \citenamefont {Watanabe}, \citenamefont {Campos}, \citenamefont {Muller},
  \citenamefont {Guo}, \citenamefont {Kim}, \citenamefont {Hone}, \citenamefont
  {Shepard},\ and\ \citenamefont {Dean}}]{high_mobility}%
  \BibitemOpen
  \bibfield  {author} {\bibinfo {author} {\bibfnamefont {L.}~\bibnamefont
  {Wang}}, \bibinfo {author} {\bibfnamefont {I.}~\bibnamefont {Meric}},
  \bibinfo {author} {\bibfnamefont {P.}~\bibnamefont {Huang}}, \bibinfo
  {author} {\bibfnamefont {Q.}~\bibnamefont {Gao}}, \bibinfo {author}
  {\bibfnamefont {Y.}~\bibnamefont {Gao}}, \bibinfo {author} {\bibfnamefont
  {H.}~\bibnamefont {Tran}}, \bibinfo {author} {\bibfnamefont {T.}~\bibnamefont
  {Taniguchi}}, \bibinfo {author} {\bibfnamefont {K.}~\bibnamefont {Watanabe}},
  \bibinfo {author} {\bibfnamefont {L.}~\bibnamefont {Campos}}, \bibinfo
  {author} {\bibfnamefont {D.}~\bibnamefont {Muller}}, \bibinfo {author}
  {\bibfnamefont {J.}~\bibnamefont {Guo}}, \bibinfo {author} {\bibfnamefont
  {P.}~\bibnamefont {Kim}}, \bibinfo {author} {\bibfnamefont {J.}~\bibnamefont
  {Hone}}, \bibinfo {author} {\bibfnamefont {K.}~\bibnamefont {Shepard}}, \
  and\ \bibinfo {author} {\bibfnamefont {C.~R.}\ \bibnamefont {Dean}},\ }\href
  {\doibase 10.1126/science.1244358} {\bibfield  {journal} {\bibinfo  {journal}
  {Science}\ }\textbf {\bibinfo {volume} {342}},\ \bibinfo {pages} {614}
  (\bibinfo {year} {2013})}\BibitemShut {NoStop}%
\bibitem [{\citenamefont {Ramezanali}\ \emph {et~al.}(2009)\citenamefont
  {Ramezanali}, \citenamefont {Vazifeh}, \citenamefont {Asgari}, \citenamefont
  {Polini},\ and\ \citenamefont {MacDonald}}]{RPA}%
  \BibitemOpen
  \bibfield  {author} {\bibinfo {author} {\bibfnamefont {M.~R.}\ \bibnamefont
  {Ramezanali}}, \bibinfo {author} {\bibfnamefont {M.~M.}\ \bibnamefont
  {Vazifeh}}, \bibinfo {author} {\bibfnamefont {R.}~\bibnamefont {Asgari}},
  \bibinfo {author} {\bibfnamefont {M.}~\bibnamefont {Polini}}, \ and\ \bibinfo
  {author} {\bibfnamefont {A.~H.}\ \bibnamefont {MacDonald}},\ }\href {\doibase
  10.1088/1751-8113/42/21/214015} {\bibfield  {journal} {\bibinfo  {journal}
  {J. Phys. A: Math. Theor.}\ }\textbf {\bibinfo {volume} {42}},\ \bibinfo
  {pages} {214015} (\bibinfo {year} {2009})}\BibitemShut {NoStop}%
\bibitem [{\citenamefont {Luo}\ \emph {et~al.}(2013)\citenamefont {Luo},
  \citenamefont {Qiu}, \citenamefont {Lu},\ and\ \citenamefont {Ni}}]{TETM}%
  \BibitemOpen
  \bibfield  {author} {\bibinfo {author} {\bibfnamefont {X.}~\bibnamefont
  {Luo}}, \bibinfo {author} {\bibfnamefont {T.}~\bibnamefont {Qiu}}, \bibinfo
  {author} {\bibfnamefont {W.}~\bibnamefont {Lu}}, \ and\ \bibinfo {author}
  {\bibfnamefont {Z.}~\bibnamefont {Ni}},\ }\href {\doibase
  10.1016/j.mser.2013.09.001} {\bibfield  {journal} {\bibinfo  {journal}
  {Mater. Sci. Eng. R Rep.}\ }\textbf {\bibinfo {volume} {74}},\ \bibinfo
  {pages} {351} (\bibinfo {year} {2013})}\BibitemShut {NoStop}%
\bibitem [{\citenamefont {Ni}\ \emph {et~al.}(2018)\citenamefont {Ni},
  \citenamefont {McLeod}, \citenamefont {Sun}, \citenamefont {Wang},
  \citenamefont {Xiong}, \citenamefont {Post}, \citenamefont {Sunku},
  \citenamefont {Jiang}, \citenamefont {Hone}, \citenamefont {Dean},
  \citenamefont {Fogler},\ and\ \citenamefont {Basov}}]{basov_nature_2018}%
  \BibitemOpen
  \bibfield  {author} {\bibinfo {author} {\bibfnamefont {G.~X.}\ \bibnamefont
  {Ni}}, \bibinfo {author} {\bibfnamefont {A.~S.}\ \bibnamefont {McLeod}},
  \bibinfo {author} {\bibfnamefont {Z.}~\bibnamefont {Sun}}, \bibinfo {author}
  {\bibfnamefont {L.}~\bibnamefont {Wang}}, \bibinfo {author} {\bibfnamefont
  {L.}~\bibnamefont {Xiong}}, \bibinfo {author} {\bibfnamefont {K.~W.}\
  \bibnamefont {Post}}, \bibinfo {author} {\bibfnamefont {S.~S.}\ \bibnamefont
  {Sunku}}, \bibinfo {author} {\bibfnamefont {B.-Y.}\ \bibnamefont {Jiang}},
  \bibinfo {author} {\bibfnamefont {J.}~\bibnamefont {Hone}}, \bibinfo {author}
  {\bibfnamefont {C.~R.}\ \bibnamefont {Dean}}, \bibinfo {author}
  {\bibfnamefont {M.~M.}\ \bibnamefont {Fogler}}, \ and\ \bibinfo {author}
  {\bibfnamefont {D.~N.}\ \bibnamefont {Basov}},\ }\href {\doibase
  10.1038/s41586-018-0136-9} {\bibfield  {journal} {\bibinfo  {journal}
  {Nature}\ }\textbf {\bibinfo {volume} {577}},\ \bibinfo {pages} {530}
  (\bibinfo {year} {2018})}\BibitemShut {NoStop}%
\bibitem [{\citenamefont {Lundeberg}\ \emph {et~al.}(2017)\citenamefont
  {Lundeberg}, \citenamefont {Gao}, \citenamefont {Asgari}, \citenamefont
  {Tan}, \citenamefont {adn Marta Autore~adn Pablo Alonso-González},
  \citenamefont {Woessner}, \citenamefont {Watanabe}, \citenamefont
  {Taniguchi}, \citenamefont {Hillenbrand}, \citenamefont {Hone}, \citenamefont
  {Polini},\ and\ \citenamefont {Koppens}}]{koppens_science_17}%
  \BibitemOpen
  \bibfield  {author} {\bibinfo {author} {\bibfnamefont {M.~B.}\ \bibnamefont
  {Lundeberg}}, \bibinfo {author} {\bibfnamefont {Y.}~\bibnamefont {Gao}},
  \bibinfo {author} {\bibfnamefont {R.}~\bibnamefont {Asgari}}, \bibinfo
  {author} {\bibfnamefont {C.}~\bibnamefont {Tan}}, \bibinfo {author}
  {\bibfnamefont {B.~V.~D.}\ \bibnamefont {adn Marta Autore~adn Pablo
  Alonso-González}}, \bibinfo {author} {\bibfnamefont {A.}~\bibnamefont
  {Woessner}}, \bibinfo {author} {\bibfnamefont {K.}~\bibnamefont {Watanabe}},
  \bibinfo {author} {\bibfnamefont {T.}~\bibnamefont {Taniguchi}}, \bibinfo
  {author} {\bibfnamefont {R.}~\bibnamefont {Hillenbrand}}, \bibinfo {author}
  {\bibfnamefont {J.}~\bibnamefont {Hone}}, \bibinfo {author} {\bibfnamefont
  {M.}~\bibnamefont {Polini}}, \ and\ \bibinfo {author} {\bibfnamefont
  {F.}~\bibnamefont {Koppens}},\ }\href {\doibase 10.1126/science.aan2735}
  {\bibfield  {journal} {\bibinfo  {journal} {Science}\ }\textbf {\bibinfo
  {volume} {357}},\ \bibinfo {pages} {187} (\bibinfo {year}
  {2017})}\BibitemShut {NoStop}%
\bibitem [{\citenamefont {Torre}\ \emph {et~al.}(2017)\citenamefont {Torre},
  \citenamefont {Katsnelson}, \citenamefont {Diaspro}, \citenamefont
  {Pellegrini},\ and\ \citenamefont {Polini}}]{chinn}%
  \BibitemOpen
  \bibfield  {author} {\bibinfo {author} {\bibfnamefont {I.}~\bibnamefont
  {Torre}}, \bibinfo {author} {\bibfnamefont {M.~I.}\ \bibnamefont
  {Katsnelson}}, \bibinfo {author} {\bibfnamefont {A.}~\bibnamefont {Diaspro}},
  \bibinfo {author} {\bibfnamefont {V.}~\bibnamefont {Pellegrini}}, \ and\
  \bibinfo {author} {\bibfnamefont {M.}~\bibnamefont {Polini}},\ }\href
  {\doibase 10.1103/PhysRevB.96.035433} {\bibfield  {journal} {\bibinfo
  {journal} {Phys. Rev. B}\ }\textbf {\bibinfo {volume} {96}},\ \bibinfo
  {pages} {035433} (\bibinfo {year} {2017})}\BibitemShut {NoStop}%
\bibitem [{\citenamefont {Xiong}\ \emph {et~al.}(2019)\citenamefont {Xiong},
  \citenamefont {Forsythe}, \citenamefont {Jung}, \citenamefont {McLeod},
  \citenamefont {Sunku}, \citenamefont {Shao}, \citenamefont {Ni},
  \citenamefont {Sternbach}, \citenamefont {Liu}, \citenamefont {Edgar},
  \citenamefont {Mele}, \citenamefont {Fogler}, \citenamefont {Shvets},
  \citenamefont {Dean},\ and\ \citenamefont {Basov}}]{PCGP}%
  \BibitemOpen
  \bibfield  {author} {\bibinfo {author} {\bibfnamefont {L.}~\bibnamefont
  {Xiong}}, \bibinfo {author} {\bibfnamefont {C.}~\bibnamefont {Forsythe}},
  \bibinfo {author} {\bibfnamefont {M.}~\bibnamefont {Jung}}, \bibinfo {author}
  {\bibfnamefont {A.~S.}\ \bibnamefont {McLeod}}, \bibinfo {author}
  {\bibfnamefont {S.~S.}\ \bibnamefont {Sunku}}, \bibinfo {author}
  {\bibfnamefont {Y.~M.}\ \bibnamefont {Shao}}, \bibinfo {author}
  {\bibfnamefont {G.~X.}\ \bibnamefont {Ni}}, \bibinfo {author} {\bibfnamefont
  {A.~J.}\ \bibnamefont {Sternbach}}, \bibinfo {author} {\bibfnamefont
  {S.}~\bibnamefont {Liu}}, \bibinfo {author} {\bibfnamefont {J.~H.}\
  \bibnamefont {Edgar}}, \bibinfo {author} {\bibfnamefont {E.~J.}\ \bibnamefont
  {Mele}}, \bibinfo {author} {\bibfnamefont {M.~M.}\ \bibnamefont {Fogler}},
  \bibinfo {author} {\bibfnamefont {G.}~\bibnamefont {Shvets}}, \bibinfo
  {author} {\bibfnamefont {C.~R.}\ \bibnamefont {Dean}}, \ and\ \bibinfo
  {author} {\bibfnamefont {D.~N.}\ \bibnamefont {Basov}},\ }\href {\doibase
  10.1038/s41467-019-12778-2} {\bibfield  {journal} {\bibinfo  {journal} {Nat.
  Commun.}\ }\textbf {\bibinfo {volume} {10}},\ \bibinfo {pages} {4780}
  (\bibinfo {year} {2019})}\BibitemShut {NoStop}%
\bibitem [{\citenamefont {Kurman}\ and\ \citenamefont
  {Kaminer}(2020)}]{Ido2020}%
  \BibitemOpen
  \bibfield  {author} {\bibinfo {author} {\bibfnamefont {Y.}~\bibnamefont
  {Kurman}}\ and\ \bibinfo {author} {\bibfnamefont {I.}~\bibnamefont
  {Kaminer}},\ }\href {\doibase 10.1038/s41567-020-0890-0} {\bibfield
  {journal} {\bibinfo  {journal} {Nat. Phys.}\ }\textbf {\bibinfo {volume}
  {16}},\ \bibinfo {pages} {868} (\bibinfo {year} {2020})}\BibitemShut
  {NoStop}%
\bibitem [{\citenamefont {Kyriienko}\ and\ \citenamefont
  {Shelykh}(2012)}]{IPP1}%
  \BibitemOpen
  \bibfield  {author} {\bibinfo {author} {\bibfnamefont {O.}~\bibnamefont
  {Kyriienko}}\ and\ \bibinfo {author} {\bibfnamefont {I.~A.}\ \bibnamefont
  {Shelykh}},\ }\href {\doibase 10.1117/1.JNP.6.061804} {\bibfield  {journal}
  {\bibinfo  {journal} {J. or Nanophotonics}\ }\textbf {\bibinfo {volume}
  {6}},\ \bibinfo {pages} {061804} (\bibinfo {year} {2012})}\BibitemShut
  {NoStop}%
\bibitem [{\citenamefont {Zanotto}\ \emph {et~al.}(2010)\citenamefont
  {Zanotto}, \citenamefont {Biasiol}, \citenamefont {Degl’Innocenti},
  \citenamefont {Sorba1},\ and\ \citenamefont {Tredicucci}}]{IPP2}%
  \BibitemOpen
  \bibfield  {author} {\bibinfo {author} {\bibfnamefont {S.}~\bibnamefont
  {Zanotto}}, \bibinfo {author} {\bibfnamefont {G.}~\bibnamefont {Biasiol}},
  \bibinfo {author} {\bibfnamefont {R.}~\bibnamefont {Degl’Innocenti}},
  \bibinfo {author} {\bibfnamefont {L.}~\bibnamefont {Sorba1}}, \ and\ \bibinfo
  {author} {\bibfnamefont {A.}~\bibnamefont {Tredicucci}},\ }\href {\doibase
  10.1063/1.3524823} {\bibfield  {journal} {\bibinfo  {journal} {Appl. Phys.
  Lett.}\ }\textbf {\bibinfo {volume} {97}},\ \bibinfo {pages} {231123}
  (\bibinfo {year} {2010})}\BibitemShut {NoStop}%
\bibitem [{\citenamefont {Anappara}\ \emph {et~al.}(2009)\citenamefont
  {Anappara}, \citenamefont {Liberato}, \citenamefont {Tredicucci},
  \citenamefont {Ciuti}, \citenamefont {Biasiol}, \citenamefont {Sorba},\ and\
  \citenamefont {Beltram1}}]{ultrastrong}%
  \BibitemOpen
  \bibfield  {author} {\bibinfo {author} {\bibfnamefont {A.~A.}\ \bibnamefont
  {Anappara}}, \bibinfo {author} {\bibfnamefont {S.~D.}\ \bibnamefont
  {Liberato}}, \bibinfo {author} {\bibfnamefont {A.}~\bibnamefont
  {Tredicucci}}, \bibinfo {author} {\bibfnamefont {C.}~\bibnamefont {Ciuti}},
  \bibinfo {author} {\bibfnamefont {G.}~\bibnamefont {Biasiol}}, \bibinfo
  {author} {\bibfnamefont {L.}~\bibnamefont {Sorba}}, \ and\ \bibinfo {author}
  {\bibfnamefont {F.}~\bibnamefont {Beltram1}},\ }\href {\doibase
  10.1103/PhysRevB.79.201303} {\bibfield  {journal} {\bibinfo  {journal} {Phys.
  Rev. B}\ }\textbf {\bibinfo {volume} {79}},\ \bibinfo {pages} {201303(R)}
  (\bibinfo {year} {2009})}\BibitemShut {NoStop}%
\bibitem [{\citenamefont {Lee}\ \emph {et~al.}(2014)\citenamefont {Lee},
  \citenamefont {Tymchenko}, \citenamefont {Argyropoulos}, \citenamefont
  {Chen}, \citenamefont {Lu}, \citenamefont {Demmerle}, \citenamefont {Boehm},
  \citenamefont {Amann}, \citenamefont {Alù},\ and\ \citenamefont
  {Belkin}}]{Lee2014}%
  \BibitemOpen
  \bibfield  {author} {\bibinfo {author} {\bibfnamefont {J.}~\bibnamefont
  {Lee}}, \bibinfo {author} {\bibfnamefont {M.}~\bibnamefont {Tymchenko}},
  \bibinfo {author} {\bibfnamefont {C.}~\bibnamefont {Argyropoulos}}, \bibinfo
  {author} {\bibfnamefont {P.-Y.}\ \bibnamefont {Chen}}, \bibinfo {author}
  {\bibfnamefont {F.}~\bibnamefont {Lu}}, \bibinfo {author} {\bibfnamefont
  {F.}~\bibnamefont {Demmerle}}, \bibinfo {author} {\bibfnamefont
  {G.}~\bibnamefont {Boehm}}, \bibinfo {author} {\bibfnamefont {M.-C.}\
  \bibnamefont {Amann}}, \bibinfo {author} {\bibfnamefont {A.}~\bibnamefont
  {Alù}}, \ and\ \bibinfo {author} {\bibfnamefont {M.~A.}\ \bibnamefont
  {Belkin}},\ }\href {\doibase 10.1038/nature13455} {\bibfield  {journal}
  {\bibinfo  {journal} {Nature}\ }\textbf {\bibinfo {volume} {511}},\ \bibinfo
  {pages} {65} (\bibinfo {year} {2014})}\BibitemShut {NoStop}%
\bibitem [{\citenamefont {kun Shi}\ \emph {et~al.}(2019)\citenamefont {kun
  Shi}, \citenamefont {Ma},\ and\ \citenamefont {Song}}]{SL_TMD}%
  \BibitemOpen
  \bibfield  {author} {\bibinfo {author} {\bibfnamefont {L.}~\bibnamefont {kun
  Shi}}, \bibinfo {author} {\bibfnamefont {J.}~\bibnamefont {Ma}}, \ and\
  \bibinfo {author} {\bibfnamefont {J.~C.~W.}\ \bibnamefont {Song}},\ }\href
  {\doibase 10.1088/2053-1583/ab59a8} {\bibfield  {journal} {\bibinfo
  {journal} {2D Mater.}\ }\textbf {\bibinfo {volume} {7}},\ \bibinfo {pages}
  {015028} (\bibinfo {year} {2019})}\BibitemShut {NoStop}%
\end{thebibliography}

%

\newpage
\title{Supplementary Materials: Emergent intersubband-plasmon-polaritons of Dirac electrons under one-dimensional superlattice
}
\maketitle
\setcounter{figure}{0}
\setcounter{equation}{0}
\renewcommand{\thetable}{S\arabic{table}}
\renewcommand{\theequation}{S\arabic{equation}}
\renewcommand{\thefigure}{S\arabic{figure}}
\section{Bound states of Dirac electrons in square potential wells}
Here, we provide an analytical solution for bound states based on total internal reflection of electrons with Dirac dispersion (TIREDD) in a square potential well or a square potential barrier. Suppose a square potential well ($U_0<U_1$) or a square potential barrier ($U_0>U_1$) given as
\begin{equation}
\label{eq.s.1.}
U_E(x) = U_0 \text{  }(0<x<W), \text{  }U_1 \text{  }(\text{elsewhere}).
\end{equation}
Then, we can set an ansatz for a bound state with eigenenergy $E$ as
\begin{equation}
\label{eq.s.2.}
\psi(\mathbf{r}) = e^{ik_y y} \times \begin{cases}
  L\begin{bmatrix}
  1 \\ \frac{-i\alpha+ik_y}{(E-U_1)/\hbar v_F}
  \end{bmatrix}e^{\alpha x}  & (x<0) \\
  A\begin{bmatrix}
  1 \\ \frac{q+ik_y}{(E-U_0)/\hbar v_F}
  \end{bmatrix}e^{i q x}+B\begin{bmatrix}
  1 \\ \frac{-q+ik_y}{(E-U_0)/\hbar v_F}
  \end{bmatrix}e^{-i q x}  & (0<x<W) \\
  R\begin{bmatrix}
  1 \\ \frac{i\alpha+ik_y}{(E-U_1)/\hbar v_F}
  \end{bmatrix}e^{-\alpha x}  & (x>W)
\end{cases},
\end{equation}
where $q=\sqrt{\left(\frac{E-U_0}{\hbar v_F}\right)^2 - k_y^2}$ is the momentum along $x$-axis within the well/barrier, and $\alpha=\sqrt{k_y^2 - \left(\frac{E-U_1}{\hbar v_F}\right)^2}$ is the decaying factor outside the well/barrier. Both $q$ and $\alpha$ are real; therefore, this ansatz is possible only when $|E-U_1|<\hbar v_F k_y<|E-U_0|$. 

\begin{figure}
    \includegraphics[width=0.7\columnwidth]{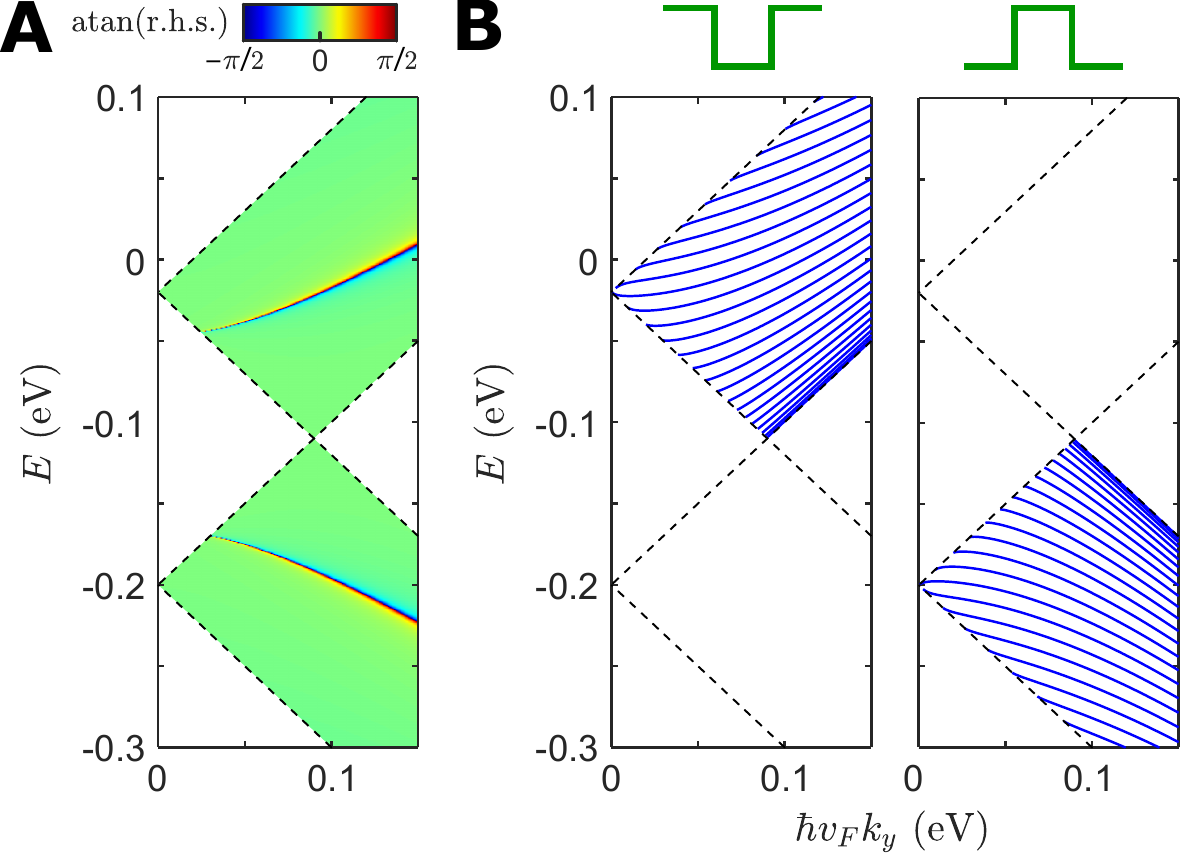}
    \centering
    \caption{\textbf{a}. Plotting the inverse tangent of the right hand side of Eq. (\ref{eq.s.3.}); $U_0=-0.2$eV, $U_1=-0.02$eV (or $U_0=-0.02$eV, $U_1=-0.2$eV; both yield the same result), and $W=200$nm ($v_F=1.1\times10^6$m/s). \textbf{b}. Dispersion of bound states in $k_y$ for a potential well (left: $U_0=-0.2$eV, $U_1=-0.02$eV) and for a potential barrier (left: $U_0=-0.02$eV, $U_1=-0.2$eV). The dased lines refer to $\hbar v_F k_y=|E-U_0|$ and $\hbar v_F k_y=|E-U_1|$.}
    \label{fig.S.1.}
\end{figure}

By imposing the continuity of $\psi$ and the continuity of probability current $\mathbf{J} = \psi^\dagger \sigma_x \psi \hat{\mathbf{x}}+\psi^\dagger \sigma_y \psi \hat{\mathbf{y}}$, we obtain the following condition for the eigenenergy $E$:
\begin{equation}
\label{eq.s.3.}
\tan(qW)=\frac{q\alpha}{\frac{E-U_0}{\hbar v_F}\frac{E-U_1}{\hbar v_F}-k_y^2}.
\end{equation}
Figure \ref{fig.S.1.}\textbf{a} shows the inverse tangent of the right hand side of the above equation. Since the right hand side is vanisihng in most region, the bound state energy condition simply reduces to $\tan(qW)\sim0$. Therefore, we get $E_j\sim\hbar v_F \sqrt{(\pi j/W)^2 + k_y^2}+U_0$, which we discussed in the main text. Figure \ref{fig.S.1.}\textbf{b} shows that a potential barrier also can host bound states via TIREDD of the valence band electrons, as discussed in the main text.

\section{Numerical method for the calculation of HIPP dispersion and normal reflection spectrum}
In this section, we elaborate on the numerical method used for the calculation of HIPP dispersion and normal reflection spectrum shown in the main text. As discussed in the main text, we only consider the transverse magnetic modes that can be described with $E_x$, $E_z$ and $B_y$. Also, we consider the dispersion in the momentum along $x$-axis ($\mathbf{q}=q\hat{\mathbf{x}}$).

Recall that the metagate periodicity is $L$, and the width of air gaps in the metagate is $S$. For convenience, let's define several variables to describe the plane wave solutions in each of the layers---the air above all layers ($A$), hBN layers ($B$), the oxide layer ($O$), and the substrate ($S$): $q_m = q+\frac{2\pi m}{L}$, $q_\omega = \frac{\omega}{c}$, $\kappa_m^A = -i\sqrt{q_\omega^2 - q_m^2}$, $\kappa_m^B = -i\sqrt{\epsilon_{xy}^B q_\omega^2 - \frac{\epsilon_{xy}^B}{\epsilon_{z}^B}q_m^2}$, $\kappa_m^O = -i\sqrt{\epsilon^O q_\omega^2 - q_m^2}$, and $\kappa_m^S = -i\sqrt{\epsilon^S q_\omega^2 - q_m^2}$, where $m\in\mathbb{Z}$ is an integer index. To describe the modes ($E_x$, $E_z$ and $B_y$) in the air gaps of the metagate ($M$): $\eta_\mu = \frac{\pi\mu}{S}$, $\phi_\mu(x)=\sqrt{2-\delta_{\mu0}}\cos\left[\eta_\mu(x-\frac{L-S}{2})\right]$, and $\kappa_\mu^M = -i\sqrt{q_\omega^2 - \eta_\mu^2}$, where $\mu$ is a non-negative integer index.

Then, we can set an ansatz for the HIPP mode as below.

(i) In the air above the top hBN layer ($z>h_t$):
\begin{equation}
\label{eq.s.4.}
\begin{split}
cB_y&=\sum_m e^{iq_mx}\left(R_m e^{-\kappa_m^A(z-h_t)} + I_{m} e^{\kappa_m^A(z-h_t)}\right)\\
k_\omega E_x&=-i\sum_m e^{iq_mx}\kappa_m^A\left(-R_m e^{-\kappa_m^A(z-h_t)} + I_{m} e^{\kappa_m^A(z-h_t)}\right)\\
k_\omega E_z&=-\sum_m e^{iq_mx}q_m\left(R_m e^{-\kappa_m^A(z-h_t)} + I_{m} e^{\kappa_m^A(z-h_t)}\right).
\end{split}
\end{equation}
Here, $R_0$ is the reflection coefficient, in the presence of the normal ($q=0$) incident lght $I_m=\delta_{m0}$. An eigenmode exists even with vanighing external drive term $I_m=0$.

(ii) In the top hBN layer ($0<z<h_t$):
\begin{equation}
\label{eq.s.5.}
\begin{split}
cB_y&=\sum_m e^{iq_mx}\left(A_m \cosh(\kappa_m^B z) + B_m \sinh(\kappa_m^B z)\right)\\
k_\omega E_x&=\frac{-i}{\epsilon_{xy}^B}\sum_m e^{iq_mx}\kappa_m^B\left(A_m \sinh(\kappa_m^B z) + B_m \cosh(\kappa_m^B z)\right)\\
k_\omega E_z&=\frac{-1}{\epsilon_{z}^B}\sum_m e^{iq_mx}q_m\left(A_m \cosh(\kappa_m^B z) + B_m \sinh(\kappa_m^B z)\right)
\end{split}
\end{equation}

(iii) In the bottom hBN layer ($-h_b<z<0$):
\begin{equation}
\label{eq.s.6.}
\begin{split}
cB_y&=\sum_m e^{iq_mx}\left(C_m \cosh(\kappa_m^B z) + B_m \sinh(\kappa_m^B z)\right)\\
k_\omega E_x&=\frac{-i}{\epsilon_{xy}^B}\sum_m e^{iq_mx}\kappa_m^B\left(C_m \sinh(\kappa_m^B z) + B_m \cosh(\kappa_m^B z)\right)\\
k_\omega E_z&=\frac{-1}{\epsilon_{z}^B}\sum_m e^{iq_mx}q_m\left(C_m \cosh(\kappa_m^B z) + B_m \sinh(\kappa_m^B z)\right)
\end{split}
\end{equation}

(iv) In the air gap of the metagate ($-h_b-h_m<z<-h_b$):
\begin{equation}
\label{eq.s.7.}
\begin{split}
cB_y&=\sum_\mu \phi_\mu(x)\left(D_\mu \cosh(\kappa_\mu^M (z+h_b)) + E_\mu \sinh(\kappa_\mu^M (z+h_b))\right)\\
k_\omega E_x&=-i\sum_\mu \phi_\mu(x)\kappa_\mu^M\left(D_\mu \sinh(\kappa_\mu^M (z+h_b)) + E_\mu \cosh(\kappa_\mu^M (z+h_b))\right)\\
k_\omega E_z&=i\sum_\mu \partial_x\phi_\mu(x) \left(D_\mu \cosh(\kappa_\mu^M (z+h_b)) + E_\mu \sinh(\kappa_\mu^M (z+h_b))\right)
\end{split}
\end{equation}

(v) In the oxide layer ($-h_b-h_m-h_o<z<-h_b-h_m$):
\begin{equation}
\label{eq.s.8.}
\begin{split}
cB_y&=\sum_m e^{iq_mx}\left(F_m \cosh(\kappa_m^O (z+h_b+h_m)) + G_m \sinh(\kappa_m^O (z+h_b+h_m))\right)\\
k_\omega E_x&=\frac{-i}{\epsilon^O}\sum_m e^{iq_mx}\kappa_m^O\left(F_m \sinh(\kappa_m^O (z+h_b+h_m)) + G_m \cosh(\kappa_m^O (z+h_b+h_m))\right)\\
k_\omega E_z&=\frac{-1}{\epsilon^O}\sum_m e^{iq_mx}q_m\left(F_m \cosh(\kappa_m^O (z+h_b+h_m)) + G_m \sinh(\kappa_m^O (z+h_b+h_m))\right)
\end{split}
\end{equation}

(vi) In the substrate ($z<-h_b-h_m-h_o$):
\begin{equation}
\label{eq.s.9.}
\begin{split}
cB_y&=\sum_m e^{iq_mx}H_m e^{\kappa_m^S(z+h_b+h_m+h_o)}\\
k_\omega E_x&=\frac{-i}{\epsilon^S}\sum_m e^{iq_mx}\kappa_m^S H_m e^{\kappa_m^S(z+h_b+h_m+h_o)}\\
k_\omega E_z&=\frac{-1}{\epsilon^S}\sum_m e^{iq_mx}q_m H_m e^{\kappa_m^S(z+h_b+h_m+h_o)}
\end{split}
\end{equation}

At $z=0$, we can define the dynamic electric potential field on graphene $\delta U_E$ from $-\partial_x(\delta U_E/(-e)) = E_x(z=0)$, and the dynamic carrier density oscillation $\delta n$ from $-e\delta n = D_z(z=0^+)-D_z(z=0^-)$
\begin{equation}
\label{eq.s.10.}
\begin{split}
k_\omega\delta U_E &= \frac{-e}{\epsilon_{xy}^B}\sum_m \frac{\kappa_m^B}{q_m}B_m e^{iq_mx}\\
k_\omega\delta n &= \frac{1}{e}\sum_m q_m (A_m-C_m) e^{iq_mx}
\end{split}
\end{equation}

Now, we match the boundary conditions (continuity of $E_x$ and continuity of $B_y$ with no free current density). At $z=0$, due to the current density at graphene, we get $B_y(z=0^+)-B_y(z=0^-) = -\mu_0 \sigma * E_x(z=0)$. Again, for convenience, let's define several vector/matrix notations: 

$\{ A_m \}, \{ B_m \}, ..., \{ I_m \}, \{ R_m \} \rightarrow A,B, ..., I, R$, 

$[\mathcal{C}_t]_{mm^\prime}=\delta_{mm^\prime}\cosh(\kappa_m^B h_t)$,  $[\mathcal{S}_t]_{mm^\prime}=\delta_{mm^\prime}\sinh(\kappa_m^B h_t)$,  

$[\mathcal{C}_b]_{mm^\prime}=\delta_{mm^\prime}\cosh(\kappa_m^B h_b)$,  $[\mathcal{S}_b]_{mm^\prime}=\delta_{mm^\prime}\sinh(\kappa_m^B h_b)$,  

$[\mathcal{C}_m]_{\mu\mu^\prime}=\delta_{\mu\mu^\prime}\cosh(\kappa_\mu^M h_m)$,  $[\mathcal{S}_m]_{\mu\mu^\prime}=\delta_{\mu\mu^\prime}\sinh(\kappa_\mu^M h_m)$,  

$[\mathcal{C}_o]_{mm^\prime}=\delta_{mm^\prime}\cosh(\kappa_m^O h_o)$,  $[\mathcal{S}_o]_{mm^\prime}=\delta_{mm^\prime}\sinh(\kappa_m^O h_o)$,  

$[\mathcal{K}^A]_{mm^\prime}=\delta_{mm^\prime}\kappa_m^A$,  $[\mathcal{K}^B]_{mm^\prime}=\delta_{mm^\prime}\kappa_m^B/\epsilon_{xy}^B$,  $[\mathcal{K}^M]_{\mu\mu^\prime}=\delta_{\mu\mu^\prime}\kappa_\mu^M$,  $[\mathcal{K}^O]_{mm^\prime}=\delta_{mm^\prime}\kappa_m^O/\epsilon^O$,  $[\mathcal{K}^S]_{mm^\prime}=\delta_{mm^\prime}\kappa_m^S/\epsilon^S$, $[\mathcal{Q}]_{mm^\prime}=\delta_{mm^\prime}q_m$,

$[\Sigma]_{mm^\prime}=-i\sigma(q_n,q_m,\omega)$, and $[\mathcal{T}]_{\mu m}=\frac{1}{S}\int_S dx \psi_\mu^*(x)e^{iq_m x}$. Here, $\mathcal{T}$ is the basis transformation matrix (from plane waves to the eigenmodes in the air gap). From the boundary conditions, we get:

(i) Between the air and the top hBN:
\begin{equation}
\label{eq.s.11.}
\begin{split}
\mathcal{C}_t A + \mathcal{S}_t B &= R + I\\
\mathcal{K}^B(\mathcal{S}_t A + \mathcal{C}_t B) &= \mathcal{K}^A(-R + I)
\end{split}
\end{equation}

(ii) Between the top and the bottom hBN:
\begin{equation}
\label{eq.s.12.}
C-A = \frac{1}{\omega} \Sigma \mathcal{K}^B B
\end{equation}

(iii) Between the bottom BN and the air gap of the metagate:
\begin{equation}
\label{eq.s.13.}
\begin{split}
D &= \mathcal{T}(\mathcal{C}_b C - \mathcal{S}_b B)\\
\mathcal{K}^M E &= \mathcal{T}\mathcal{K}^B (\mathcal{C}_b B - \mathcal{S}_b C)\\
\frac{S}{L}\mathcal{T}^\dagger \mathcal{K}^M E &= \mathcal{K}^B (\mathcal{C}_b B - \mathcal{S}_b C)
\end{split}
\end{equation}

(iv) Between the air gap of the metagate and the oxide:
\begin{equation}
\label{eq.s.14.}
\begin{split}
\mathcal{C}_m D - \mathcal{S}_m E&= \mathcal{T}F\\
\mathcal{K}^M (\mathcal{C}_m E - \mathcal{S}_m D) &= \mathcal{T}\mathcal{K}^O G\\
\frac{S}{L}\mathcal{T}^\dagger \mathcal{K}^M (\mathcal{C}_m E - \mathcal{S}_m D) &= \mathcal{K}^O G
\end{split}
\end{equation}

(v) Between the oxide and the substrate:
\begin{equation}
\label{eq.s.15.}
\begin{split}
\mathcal{C}_o F - \mathcal{S}_o G&= H\\
\mathcal{K}^O(\mathcal{C}_o G - \mathcal{S}_o F)&= \mathcal{K}^S H
\end{split}
\end{equation}

From Eq. (\ref{eq.s.15.}), we can eliminate $H$ to obtain
\begin{equation}
\label{eq.s.16.}
G = (\mathcal{K}^O\mathcal{C}_o+\mathcal{K}^S\mathcal{S}_o)^{-1}(\mathcal{K}^O\mathcal{S}_o+\mathcal{K}^S\mathcal{C}_o)F.
\end{equation}
Then, from Eq. (\ref{eq.s.14.}) and Eq. (\ref{eq.s.16.}), we can eliminate $F$ and $G$ to obtain:
\begin{equation}
\label{eq.s.17.}
E = (\mathcal{V}\mathcal{K}^M\mathcal{C}_m+\mathcal{S}_m)^{-1}(\mathcal{V}\mathcal{K}^M\mathcal{S}_m+\mathcal{C}_m)D,
\end{equation}
where $\mathcal{V} = \frac{S}{L} \mathcal{T}(\mathcal{K}^O\mathcal{S}_o+\mathcal{K}^S\mathcal{C}_o)^{-1}(\mathcal{K}^O\mathcal{C}_o+\mathcal{K}^S\mathcal{S}_o){K^O}^{-1}\mathcal{T}^\dagger$. Further, from Eq. (\ref{eq.s.13.}) and Eq. (\ref{eq.s.17.}), we can eliminate $D$ and $E$ to obtain:
\begin{equation}
\label{eq.s.18.}
C = (\mathcal{W}\mathcal{C}_b+\mathcal{K}^B\mathcal{S}_b)^{-1}(\mathcal{W}\mathcal{S}_b+\mathcal{K}^B\mathcal{C}_b)B = \left[{\mathcal{K}^B}^{-1}\mathcal{S}_b{\mathcal{C}_b}^{-1} +(\mathcal{C}_b \mathcal{W} \mathcal{C}_b + \mathcal{K}^B\mathcal{C}_b\mathcal{S}_b)^{-1}\right]\mathcal{K}^B B,
\end{equation}
where $\mathcal{W} = \frac{S}{L} \mathcal{T}^\dagger\mathcal{K}^M(\mathcal{V}\mathcal{K}^M\mathcal{C}_m+\mathcal{S}_m)^{-1}(\mathcal{V}\mathcal{K}^M\mathcal{S}_m+\mathcal{C}_m)\mathcal{T}$ or, equivalently, $\mathcal{W} = \frac{S}{L} \mathcal{T}^\dagger\left[\mathcal{K}^M\mathcal{S}_m{\mathcal{C}_m}^{-1} +(\mathcal{C}_b \mathcal{V} \mathcal{C}_b + {\mathcal{K}^M}^{-1}\mathcal{C}_m\mathcal{S}_m)^{-1}\right]\mathcal{T}$ (the latter form makes it clear that $\mathcal{W}$ is a Hermitian matrix). From Eq. (\ref{eq.s.11.}), we can eliminate eliminate $R$ to obtain:
\begin{equation}
\label{eq.s.19.}
A = 2\mathcal{K}^A(\mathcal{K}^A\mathcal{C}_t+\mathcal{K}^B\mathcal{S}_t)^{-1}I- \left[{\mathcal{K}^B}^{-1}\mathcal{S}_t{\mathcal{C}_t}^{-1} +(\mathcal{C}_t \mathcal{K}^A \mathcal{C}_t + \mathcal{K}^B\mathcal{C}_t\mathcal{S}_t)^{-1}\right]\mathcal{K}^B B,
\end{equation}
or we can eliminate $A$ to obtain:
\begin{equation}
\label{eq.s.20.}
R = (\mathcal{K}^A\mathcal{C}_t+\mathcal{K}^B\mathcal{S}_t)^{-1}\left[(\mathcal{K}^A\mathcal{C}_t-\mathcal{K}^B\mathcal{S}_t)I-\mathcal{K}^B B\right].
\end{equation}

If there is no external drive input $I=0$, Eq. (\ref{eq.s.18.}), and Eq. (\ref{eq.s.19.}) altogether gives:
\begin{equation}
\label{eq.s.21.}
\begin{split}
C-A &= \mathcal{Y}\mathcal{K}^B B\text{,  where}\\
\mathcal{Y}=\left[{\mathcal{K}^B}^{-1}\mathcal{S}_t{\mathcal{C}_t}^{-1} +(\mathcal{C}_t \mathcal{K}^A \mathcal{C}_t + \mathcal{K}^B\mathcal{C}_t\mathcal{S}_t)^{-1}\right] &+ \left[{\mathcal{K}^B}^{-1}\mathcal{S}_b{\mathcal{C}_b}^{-1} +(\mathcal{C}_b \mathcal{W} \mathcal{C}_b + \mathcal{K}^B\mathcal{C}_b\mathcal{S}_b)^{-1}\right].
\end{split}
\end{equation}
Then, by examining Eq. (\ref{eq.s.10.}) and recalling the definition of the dynamic capacitance $\delta U_E(\mathbf{q}) = e^2\sum_{\mathbf{q}^\prime} C^{-1}(\mathbf{q},\mathbf{q}^\prime) \delta n(\mathbf{q}^\prime)$, we arrive at
\begin{equation}
\label{eq.s.22.}
C^{-1}(q_m,q_{m^\prime}) = \frac{[\mathcal{Y}^{-1}]_{m m^\prime}}{q_m q_{m^\prime}} = [\mathcal{Q}^{-1}\mathcal{Y}^{-1}\mathcal{Q}^{-1}]_{m m^\prime}.
\end{equation}
The inverse of dynamic capacitance is equal to the dynamic coulomb interaction \cite{Minwoo_PRL, chinn, PCGP}. It is easily checked that, in a suspended graphene in vacuum ($h_t, h_b \rightarrow \infty$, $\epsilon_{xy}^{B}=\epsilon_{z}^{B}=\epsilon_0$), the above expression reduces to the expression obtained by taking the 2D Fourier transform of Coulomb potential $e^2C^{-1}(q_m,q_{m^\prime})\rightarrow \delta_{mm^\prime} \frac{e^2}{2\epsilon_0 |q_m|}$. Thus, the dynamical dielectric function given in the main text is reduced as a compact matrix form:
\begin{equation}
\label{eq.s.23.}
\epsilon(\mathbf{q},\omega) = \mathbf{1} - \frac{1}{\omega} \mathcal{Q}^{-1}\mathcal{Y}^{-1}\Sigma\mathcal{Q} = \mathcal{Q}^{-1}\left[\mathbf{1} - \frac{1}{\omega}\mathcal{Y}^{-1}\Sigma\right]\mathcal{Q}.
\end{equation}
Therefore, the density of state, which is approximated as the imaginary part of the inverse of the dynamical dielectric function, is given as
\begin{equation}
\label{eq.s.24.}
DOS(\mathbf{q},\omega) = -\text{Im}\left[\text{Tr}\left([\epsilon(\mathbf{q},\omega)]^{-1}\right)\right]= -\text{Im}\left[\text{Tr}\left(\left[\mathbf{1} - \frac{1}{\omega}\mathcal{Y}^{-1}\Sigma\right]^{-1}\right)\right].
\end{equation}
This is the quantity plotted in the main text for the figure containing the HIPP dispersions.

Now, we can combine Eq. (\ref{eq.s.12.}), Eq. (\ref{eq.s.18.}), and Eq. (\ref{eq.s.19.}) to obtain:
\begin{equation}
\label{eq.s.25.}
R = (\mathcal{K}^A\mathcal{C}_t+\mathcal{K}^B\mathcal{S}_t)^{-1}\left[(\mathcal{K}^A\mathcal{C}_t-\mathcal{K}^B\mathcal{S}_t)-2\left(\mathcal{Y} - \frac{1}{\omega}\Sigma\right)^{-1}\mathcal{K}^A(\mathcal{K}^A\mathcal{C}_t+\mathcal{K}^B\mathcal{S}_t)^{-1}\right]I.
\end{equation}
As mentioned earlier, the reflection upon normal incidence is calculated as $|R_0|^2$ with $I_m=\delta_{m0}$ and $q=0$.

\begin{figure}[b]
    \includegraphics[width=0.95\columnwidth]{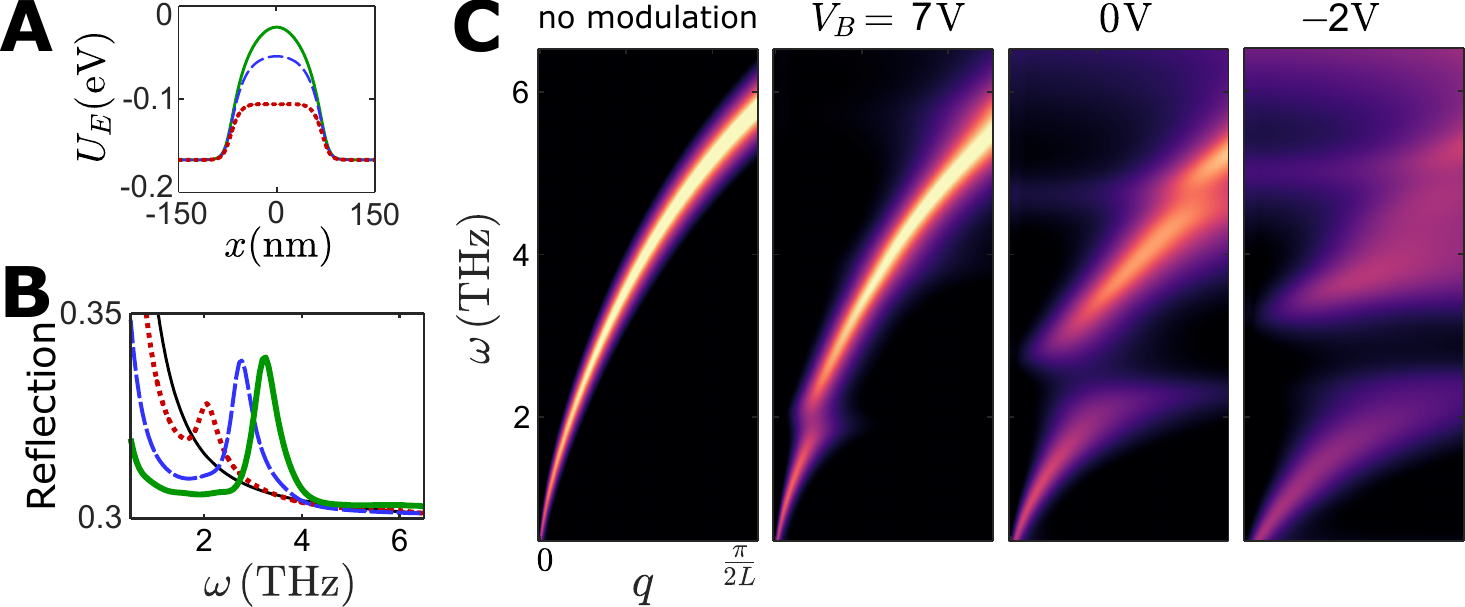}
    \centering
    \caption{\textbf{a} SL potential $U_E(x)$ (bottom) for different backgate voltages ($V_M=1$V is for all three cases); dotted red: $V_B=7$V, dashed blue: $0$V, and solid green: $-2$V. ($h_t=5$nm, $h_b=10$nm, $h_g=10$nm, $h_0=150$nm, $L=300$nm, $S=150$nm). \textbf{b} Reflection spectra for the normal incidence of light polarized along $x$-axis; solid thin black: no modulation ($E_F=0.15$eV), dotted red: $V_B=7$V, dashed blue: $0$V, and solid thick green: $-2$V. \textbf{c} Density of states or $-\text{Im}\left[\text{Tr}\left([\epsilon(\mathbf{q},\omega)]^{-1}\right)\right]$ for visualizing the HIPP dispersion. }
    \label{fig.S.2.}
\end{figure}

\section{HIPP dispersions under a different superlattice design}
In this section, we show that the appearance of the hybrid intersubband-plasmon-polaritons (HIPPs) shown in the main text is not contingent upon a specific set of parameter conditions, by providing the HIPP dispersions for a different superlattice design. In the main text, the results were shown for a system with a periodicity of $L=300$nm and the air gap width of $S=80$nm. Here, in Fig. \ref{fig.S.2.}, we provided the same calculations for $S=150$nm. For a similar degree of $U_E$ modulation, the intersubband transition (ISBT) frequencies are slightly blue-shifted, compared to the results in the main text, since the potential well width $L-S$ is now deceased. Other than such small details, the HIPP phenomenon is qualitatively the same. Therefore, the experimental verification of this HIPP emergence under 1D SL in graphene would be universally possible for nearly any choice of parameter conditions.

\end{document}